\pdfoutput=1

\documentclass[preprint,5p,times,twocolumn,fleqn]{elsarticle}   

\journal{Int. J. Microgravity Sci. Appl.}

\usepackage{hyperref}
\hypersetup{%
  colorlinks=true,
  pdfborder={0 0 0 },
}

\usepackage{nameref}
\usepackage{natbib}
\usepackage[separate-uncertainty=true,multi-part-units=single]{siunitx}
\usepackage{xcolor}
\usepackage{amsmath}
\usepackage[capitalize]{cleveref}
\crefformat{appendix}{#2#1#3}

\usepackage{caption}
\usepackage{xspace}
\usepackage{multirow}

\usepackage{balance}
\usepackage{longtable}

\usepackage{tabularx}
\newcolumntype{C}{>{\centering\arraybackslash}X}   
\newcolumntype{R}{>{\raggedleft\arraybackslash}X}  
\newcolumntype{L}{>{\raggedright\arraybackslash}X} 

\usepackage[version=3]{mhchem}

\newcommand{\pdif}[2]{\frac{\partial #1}{\partial #2}}

\newcommand{\bol}[1]{\mbox{\boldmath $#1$}}

\newcommand{\Rey}{\text{Re}}
\newcommand{\PR}{\text{Pr}}
\newcommand{\Pm}{\text{Pm}}
\newcommand{\Ga}{\text{Ga}}
\newcommand{\Ec}{\text{Ec}}
\newcommand{\Sp}{\text{Sp}}
\newcommand{\Bi}{\text{Bi}}
\newcommand{\Mg}{\text{Mg}}
\newcommand{\Ma}{\text{Ma}}
\newcommand{\Pl}{\text{Pl}}
\newcommand{\La}{\text{La}}
\newcommand{\Nu}{\text{Nu}}
\newcommand{\nablaHat}{\widehat{\nabla}}
\newcommand{\TiSixFour}{Ti-6Al-4V}
\newcommand{\toprule}{\hline\hline}
\newcommand{\midrule}{\hline\hline}
\newcommand{\bottomrule}{\hline\hline}

\setcitestyle{super}

\newcounter{daggerfootnote}

\begin{document}

\begin{frontmatter}

\title{Surrogate models for the magnitude of convection in droplets levitated through EML, ADL, and ESL methods}

\author[TCUG]{Takuro Usui}
\author[TCUB]{Suguru Shiratori\corref{cor}} \ead{sshrator@tcu.ac.jp}
\author[TCUG]{Kohei Tanimoto\corref{}}
\author[CIT]{Shumpei Ozawa}
\author[JAXA]{Takehiko Ishikawa}
\author[WU]{Shinsuke Suzuki}
\author[TCUB]{Hideaki Nagano}
\author[TCUB]{Kenjiro Shimano}

\cortext[cor]{Corresponding author}

\affiliation[TCUG]{organization={Graduate School of Integrative Science and Engineering, Tokyo City University},  city={Tokyo}, country={Japan}}
\affiliation[TCUB]{organization={Department of Mechanical Systems Engineering, Tokyo City University},  city={Tokyo}, country={Japan}}
\affiliation[CIT]{organization={Department of Advanced Materials Science and Engineering, Chiba Institute of Technology}, city={Chiba}, country={Japan}}
\affiliation[JAXA]{organization={Institute of Space and Astronautical Science, Japan Aerospace Exploration Agency}, city={Tsukuba, Ibaraki}, country={Japan}}
\affiliation[WU]{organization={Department of Materials Science, Department of Applied Mechanics and Aerospace Engineering, %
Kagami Memorial Research Institute for Materials Science and Technology, Waseda University}, city={Tokyo}, country={Japan}}

\begin{abstract}
Fluid flow and heat transfer in levitated droplets were numerically investigated.
Three levitation methods: 
electro-magnetic levitation (EML), 
aerodynamic levitation (ADL), 
and electro-static levitation (ESL) were considered,
and conservative laws of mass, momentum, and energy were applied as common models.
The Marangoni effect was applied as a velocity boundary condition, 
whereas heat transfer and radiation heat loss were considered as thermal boundary conditions.
As specific models to EML, the Lorentz force and Joule heat were calculated based on the analytical solution of the electromagnetic field.
For ADL model, besides the Marangoni effect, the flow driven by the surface shear force was considered. 
For ADL and ESL models, the effect of laser heating was introduced as a boundary condition.
All the equations were nondimensionalized using common scales for all three levitations.
Numerical simulations were performed for several materials and droplet sizes, 
and the results were evaluated in terms of the Reynolds number based on the maximum velocity of the flow in the droplet.
The order of magnitude of Reynolds numbers was evaluated as 
$\Rey \sim 10^4$ for EML, 
$\Rey \sim 10^3$ for ADL, and 
$\Rey \sim 10^1$ for ESL.
Based on the simulation results, we proposed simple formulas for predicting the Reynolds number of droplet internal convection
using combinations of nondimensional numbers determined from physical properties of the material and the driving conditions.
The proposed formulas can be used as surrogate models to predict the Reynolds numbers,
even for materials other than those used in this study.
\end{abstract}



\end{frontmatter}


\section{Introduction}\label{sec-Intro}
\subsection{Droplet levitation and internal flows}
Liquid droplets can be levitated by applying an external force balanced by the weight of the droplet.
Levitation methods provide many technological and scientific advantages 
in containerless material processing and measurements of thermophysical properties of molten metals~\cite{Lee2021}.
In the absence of a crucible, the risk of sample contamination is eliminated;
thus, thermophysical properties can be measured accurately.
In addition, the absence of a crucible suppresses heterogeneous nucleation by the container walls;
thus, it enables the generation of new materials of metastable phases, which can be applied to high-performance magnets~\cite{Kuribayashi2020,Hayasaka2021}.

In such applications of levitation methods, 
an external force and an applied heat source drives the convection inside the droplets, 
which may change the behavior of surface oscillations and the solidification process.
For instance in EML, 
owing to the internal convection driven by the electromagnetic force, 
the mode of the surface oscillation becomes different from the~\citet{Rayleigh1879}'s solution~\cite{Bojarevics2009}.
For this oscillation mode in EML, \citet{Cummings1991} analytically investigated the fluid flow and derived the relation between frequencies and the surface tension. 
In their analysis, the liquid surface was assumed to be approximately spherical, and higher order deviations were neglected to obtain an analytical solution. 
The validity of their assumption and the accuracy of the surface tension determined by their equation
depend on the magnitude of the internal flow, which is difficult to evaluate. 

The internal convection in a levitated droplet also affects the solidification behavior~\cite{Matson2022}.
In this context, we mention a 
{\em Hetero-3D} project\footnote[2]{Hetero-3D Project Web site: \url{https://humans-in-space.jaxa.jp/kibouser/subject/science/70412.html}}.
This project focuses on the solidification behavior of titanium alloy \TiSixFour, 
which is one of the most widely used alloys applied
as raw materials for metal additive manufacturing.
During a casting process,
grains of \TiSixFour are grown into coarse anisotropic columnar structures,
which are unfavorable in many applications.
The addition of heterogeneous nuclei effectively generates a fine isotropic grain structure~\cite{McCartney1989}.
For \TiSixFour, several grain refiners have been identified~\cite{TedmanJones2019} and
the TiC was found to be an effective grain refiner~\cite{Watanabe2020}.
The effect of TiC addition on \TiSixFour has been
experimentally investigated in directional solidification~\cite{Yamamoto2019,Date2021}.
However, in a quantitative aspect, the effect of the TiC on grain refinement has not been clarified sufficiently.
One of the indistinct issues in grain refinement is the effect of convection during solidification.
The amount of refined grains not only dependent on the amount of TiC but also on the magnitude of the convection.
If the convection can be avoided or suppressed, the pure effect of TiC addition on grain refinement can be clarified.
To clarify this expectation, solidification experiments will be conducted during the {\em Hetero-3D} project,
using the Electrostatic Levitation Furnace (ELF)~\cite{Tamaru2018,Ishikawa2022}
on board the Japanese Experiment Module ``Kibo'' of the International space station (ISS).
As a part of this project, \citet{Hanada2023} investigated the experimental preparation process 
to prevent bubble formation which can be an obstacle to observing the nucleation behavior of the samples.
This study was motivated under the Hetero-3D project, to clarify how strong convection is driven in ISS-ELF condition.
In the ISS-ELF, the droplet internal convection is also driven by the Marangoni effect due to laser heating.
The magnitude of the internal convection should be known before the experiment is conducted.

As described, information on the droplet internal convection is important for planning experiments on levitation methods.
Because the internal flows can hardly be visualized, numerical simulations are performed.

\subsection{Previous researches on droplet internal flows}
Several studies have reported the numerical modeling of the convection in the levitated droplets.
%
%
For EML method,
Bojarevics and his research group constructed a detailed thermofluidics model in an EML-levitated droplet, 
considering dynamic surface oscillations and turbulence~\cite{Bojarevics2000,Bojarevics2003,Bojarevics2009}. 
Because their model is sophisticated, its implementation may require a huge effort.
\citet{Berry2000} modeled the effect of turbulence,        
and \citet{Hyers2003} investigated the transition from laminar to turbulent flows.        
\citet{Tsukada2009}      
modeled the static magnetic field and investigated its effect on thermal conductivity measurements.
\citet{Spitans2013,Spitans2016} numerically investigated the dynamics of the free surface of EML-levitated droplets.
%
For the ADL method,
previous numerical studies are limited, compared to experimental studies.
\citet{Guo2019}          
conducted volume-of-fluid (VOF) simulation of aerodynamically levitated droplets
for the design study of experimental systems.
%
For the ESL method, 
\citet{Song2000}         
formulated the electric, thermal, and fluid flow fields for ESL system.
In their model, the static surface deformation was considered.
\citet{Huo2004}          
considered the dynamic surface deformation through Marangoni convection.
%

All the mentioned previous numerical simulations focused on a single levitation method,
and the results of these studies are limited to certain specific levitation conditions.
This situation is unfavorable from the viewpoint of experimental planning
because it is hard to compare the magnitude of droplet internal convection for different levitation methods.
\citet{Hyers2004} and \citet{Hyers2005}
conducted computational fluid dynamics (CFD) simulations for both the EML and ESL,
and showed the ranges of Reynolds numbers of the internal flow for the case of microgravity and terrestrial conditions. 
Their study provided a new perspective that showed a range of fluid flow for different levitation methods.
However, their investigations were limited to two types of levitation (EML and ESL): the ADL was not involved.
In addition, they only provided information on the range of the Reynolds number.
For experimental planning with limited opportunities, a more specific Reynolds number is preferred.
In this light, a simple formula must be constructed to predict the magnitude of the droplet internal convection.
\citet{Gao2016}          
proposed a simple nondimensional formula which predicts the levitation force of EML from physical properties, power input, and coil design. 
Their formula is useful for predicting the levitation force;
however, the magnitude of the convection cannot be directly predicted.
\citet{Xiao2019,Baker2020}        
proposed a surrogate model for convection in electromagnetically levitated droplets.
In their model, the maximum velocity and maximum shear rate are expressed 
by simple polynomials with a heating control voltage, density, viscosity, and electrical conductivity.
Such prediction through a simple formula is useful for experimental planning.
However, their surrogate model was only constructed for EML.

\subsection{Aim of the present study}
This study proposes a modified prediction model of droplet internal convection.
With the aim of providing a useful tool for planning experiments using levitation systems, we propose the following two methods:
\begin{itemize}
\item mathematical formulations of thermofluidics for three levitation systems: EML, ADL, and ESL,
\item simple formulas for predicting the Reynolds number of the droplet internal flow using the nondimensional numbers determined from 
the physical properties of materials, droplet sizes, and driving conditions.
\end{itemize}
%
For the mathematical models, 
because we want to predict the magnitude of the flow, 
we formulate the models with minimal components by applying some simplifications and assumptions.
Formulated models are implemented using a finite volume method on the open-source CFD solver OpenFOAM.
The numerical simulations were performed for several materials and different droplet sizes.
From the numerical results, we propose simple formulas to predict the Reynolds numbers using the combinations of nondimensional numbers 
that can be determined from the physical properties and droplet sizes of materials, as well as levitation conditions.

\section{Problem formulation}\label{sec-formulation}
\subsection{Overview}
For all three levitation systems, all the conservation laws and most of the boundary conditions can be commonly applied.
In the following, the common governing equations are described first. 
Then the models specific to individual levitation systems are formulated.

\subsection{Common governing equations}
For all levitation systems, 
the fluid is assumed to be an incompressible Newtonian fluid of density $\rho$, viscosity $\mu$, specific heat $c_p$, and thermal conductivity $\lambda$.
The flow is governed by the conservation of mass, momentum, and energy
\begin{subequations}
\begin{gather}
 \nabla\cdot \bol{u} = 0,      \label{eq-continuity} \\
 \pdif{\left( \rho \bol{u} \right)}{t} + \nabla \cdot \left( \rho\bol{u}\bol{u} \right)  = -\nabla  p + \mu \nabla^2 \bol{u} + \rho\bol{g} + \bol{f}_m,  \label{eq-NS} \\
 \pdif{\left( \rho c_p T \right) }{t} + \nabla \cdot \left( \rho c_p \bol{u} T \right) = \lambda \nabla^2 T + q_m,   \label{eq-energy}
\end{gather}
\end{subequations}
where
$\bol{u}$, $t$, $p$, and $T$ are field variables for velocity, time, pressure, and temperature, respectively.
$\bol{f}_m$ and $q_m$ are terms for the Lorenz force and Joule heat generation, respectively.
$\bol{g}$ denotes the vector of gravity acceleration.
For the velocity boundary condition on the free surface, 
the following Marangoni effect is applied
\begin{equation}
\mu \left( \nabla\bol{u} + \nabla\bol{u}^T \right) \cdot \bol{n} = \sigma_T \left( \bol{I} - \bol{n}\bol{n} \right) \cdot \nabla T,  \label{eq-Marangoni}
\end{equation}
where $\bol{I}$ is the identity tensor 
and operator $\left( \bol{I} - \bol{n}\bol{n} \right)$ represents an orthogonal projection of a vector onto the tangent plane determined by interface normal vector $\bol{n}$.
For the thermal boundary condition, the following heat fluxes are applied
\begin{equation}
-\lambda \nabla T \cdot \bol{n} =
  h \left(T - T_a \right)
+ \sigma_\text{SB} \varepsilon \left( T^4 - T_a^4 \right)
+ I_0  W \left( \bol{x} \right).
\end{equation}
The terms on the right-hand side are 
convective heat transfer, 
radiative heat loss, 
and heat gain by a heating laser.
$\sigma_\text{SB}$ is the Stefan-Boltzmann constant,
$T_a$ is the ambient temperature,
$\varepsilon$ is the emissivity,
and $I_0$ is the output power of the laser heat source.
The function $W \left( \bol{x} \right)$ is the spatial distribution of a laser heat source defined as 
\begin{equation}
 W \left( \bol{x} \right) = \frac{1}{2\pi R_L^2}\exp\left( -\frac{s^2}{2R_L^2} \right),   \label{eq-laserW}
\end{equation}
where $R_L$ is the radius of the laser spot
and $s$ is the orthogonal distance from the axis center of the laser spot.
Heat generated by the laser cannot always be treated as a boundary condition, 
and it may have a depth-wise distribution.
The radiant flux of the laser light can be expressed by exponential attenuation 
as $I(z) = I_0 \exp(-\eta z)$, which is known as the Lambert-Beer law.
$\eta$ is the attenuation coefficient with units of \si{1/m}.
For all the materials selected in this study, the inverse of the attenuation coefficient is much smaller than the computational grid size, 
which means all the laser power is absorbed within a single mesh. 
Therefore, the treatment of the laser heat generation as a boundary condition can be considered reasonable.

All through the models in this study, the surface shape of the liquid droplet is assumed to be spherical.
Concerning the dimensions of the spatial domain, an axisymmetric field is assumed for the EML and ADL systems, 
whereas a three-dimensional field is considered for the ESL system. 
The additional models specific to individual levitation systems are described in the following sections.

\subsection{Specific model for EML} \label{sec-EML_model}
\begin{figure*}
\centering\includegraphics[width=0.8\textwidth]{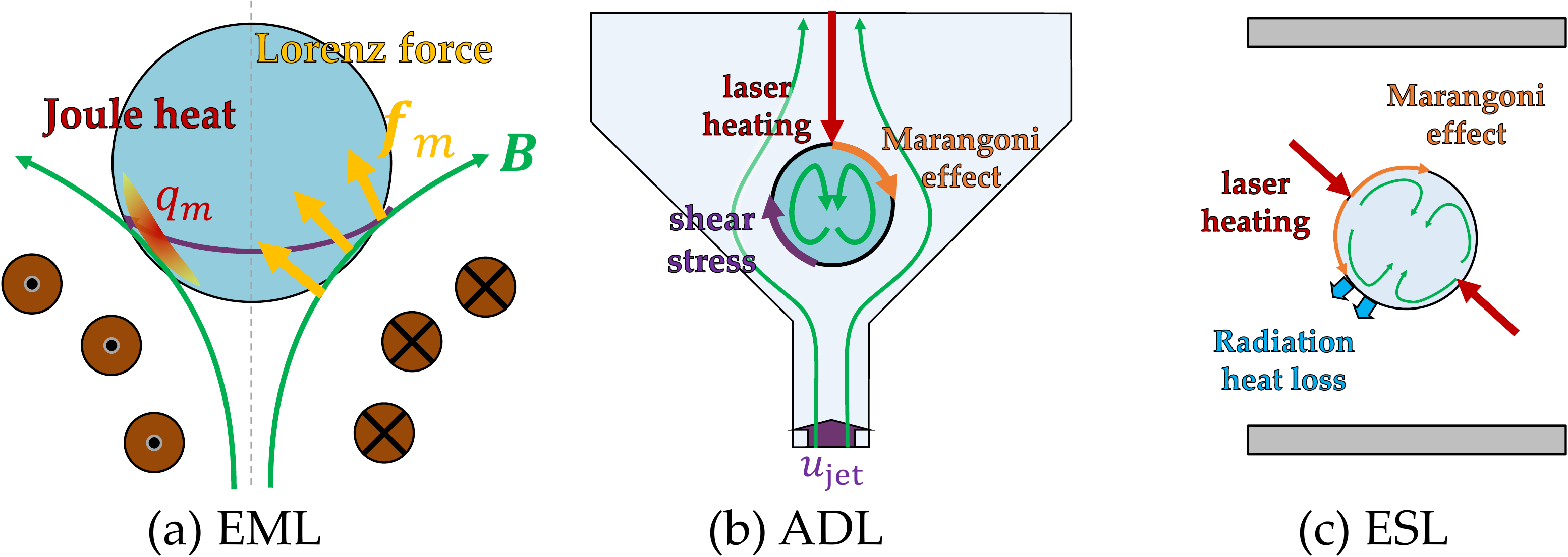}
\caption{\label{fig-All_systems} 
Schematic models for (a) EML, (b) ADL, and (c) ESL systems.}
\end{figure*}

In the model for the EML system, as shown in \cref{fig-All_systems}(a),
the Lorentz force and Joule heat must be applied
as $\bol{f}_m$ in \cref{eq-NS} and $q$ in \cref{eq-energy}, respectively.
In this study,
the electromagnetic field was formulated according to~\citet{Bojarevics2000}.
Because the thermal and flow fields in the EML system are assumed to be axisymmetric,
the symmetry property can be also applied to the electromagnetic field.
The electromagnetic field can be expressed by Faraday's and Amp\`{e}re's laws, stated as follows:
\begin{gather}
\nabla \times \bol{J} =  -\sigma_e \pdif{\bol{B}}{t},  \label{eq-Faraday} \\
\nabla \times \bol{B} =  \mu_0 \bol{J} ,               \label{eq-Ampere}
\end{gather}
where 
$\bol{B}$ is the magnetic flux density,
$\bol{J}$ is the electric current density,
$\sigma_e$ is the electrical conductivity,
and $\mu_0$ is the permeability of free space.
In \cref{eq-Faraday,eq-Ampere}, 
Ohm's law $\bol{J} = \sigma_e \bol{E}$ 
and the constitutive relation $\bol{H} = \bol{B} / \mu_0$ are assumed ($\bol{H}$ is the magnetic field strength).
In addition, the displacement current $\partial \bol{E} / \partial t$ is neglected in \cref{eq-Ampere}%
\footnote[3]{%
Generally, when considering the movement of conducting materials in the elecromagnetic filed,
the generalized Ohm's law $\bol{J} = \sigma_e \left( \bol{E} + \bol{u}\times\bol{B}  \right)$, which involves $\bol{u}\times\bol{B}$ term, is applied.
Using the scales shown in \cref{tab-scales}, nondimensional form of this equation can be written as 
$\hat{\bol{J}} = \Sp \hat{\bol{E}} + \Pm \left(\bol{U}\times\hat{\bol{B}}  \right)$.
In this study, 
nondimensional numbers $\Sp$ and $\Pm$ are evaluated as shown in \cref{tab-nondimNumbers},
where the magnitude relation
$\Pm \sim 10^{-7} \ll \Sp \sim 10^1$ 
can be clearly confirmed.
From this reason, the $\bol{u}\times\bol{B}$ term can be neglected.
}.
Under such conditions,
the electromagnetic field can be expressed using magnetic vector potential $\bol{A}$ as
\begin{gather}
\bol{B} = \nabla \times \bol{A},          \label{eq-BA} \\
\bol{J} = - \sigma_e \pdif{\bol{A}}{t}.   \label{eq-JA}
\end{gather}
For the alternate current (AC) case with angular frequency $\omega$, 
the magnetic vector potential can be expressed as $\bol{A}(\bol{x},t) = \bol{A}_0 (\bol{x}) \exp(i\omega t)$
and its time derivative is $\partial \bol{A} / \partial t = i \omega \bol{A}$.
Thus, \cref{eq-JA} becomes $\bol{J} = -i \omega \sigma_e \bol{A}$.
By substituting this into \cref{eq-BA}, the governing equation for $\bol{A}_0$ is obtained as
\begin{equation}
 \nabla^{2}  \bol{A}_0 =  i \omega \mu_0 \sigma_e \bol{A}_0.   \label{eq-govA}
\end{equation}
The analytical solution for the \cref{eq-govA} was obtained by~\citet{Smythe1968} in the spherical coordinates $(r,\theta,\varphi)$.
Under the axisymmetry assumption, 
only the azimuthal component $A_{\varphi}$ of the amplitude of magnetic vector potential $\bol{A}_0$ survives 
and \cref{eq-govA} reduces to the scalar equation for $A_{\varphi}$.
Now we consider a situation where
a sphere of radius $R_0 (=d/2)$ and electrical conductivity $\sigma_e$ is surrounded
by a current filament carrying a current of amplitude $I_s$ and angular frequency $\omega$ 
is placed at the position $r=R_s$ and $\theta = \theta_s$.
Under this condition, the analytical solution for $A_{\varphi}(r,\theta)$ can be expressed by
\begin{subequations}
\label{eq-solA}
\begin{align}
 A_{\varphi}(R, \theta)  &= \frac{\mu_0 I_s \sin\ \theta_s }{2\sqrt{i \sigma_e \mu_0 \omega R R_0}} \notag \\
& \times 
 \sum_{n=1}^{\infty} C_n{I_{n+\frac{1}{2}}  \left(R \sqrt{i\sigma_e \mu_0 \omega} \right)  P_n^{1}}  \left( \cos \theta \right),  \label{eq-solA}  \\
 C_n &= \frac{2n+1}{n(n+1)} \left( \frac{R_0}{R_s} \right)^n \frac{P_n^1 \left( \cos \theta_s \right)}{I_{n-\frac{1}{2}} \left( R_0 \sqrt{i\sigma_e \mu_0 \omega} \right) }, 
\end{align}
\end{subequations}
where $P_n^m(x)$ are the associated  Legendre polynomials,
and $I_{n + \frac{1}{2}}$ is the half-integer order modified Bessel function of complex argument.
For a detailed derication, see \citet{Li1993}.
%
After the magnetic potential $\bol{A}$ is obtained by truncating the summation in \cref{eq-solA},
the Lorentz force and Joule heat can be calculated as
\begin{gather}
\bol{f}_m = \bol{J} \times \bol{B}, \\
q_m = \frac{\left| \bol{J} \right|^2 } {\sigma_e}.
\end{gather}

\subsection{Specific model for ADL} \label{sec-model_ADL}
A model of the ADL system is schematically shown in \cref{fig-All_systems}(b).
The gas-jet flow is considered by applying velocity $u_\text{jet}$ at the boundary corresponding to the nozzle outlet.
The droplet is heated by a laser from the upper side.
In the droplet, the convection can be driven by two types of forces:
the Marangoni effect due to laser heating,
and the shear force acting on the liquid surface.
In this study, these two effects are modeled separately.
The Marangoni effect and laser heat source are already formulated in \cref{eq-Marangoni} and \cref{eq-laserW}.
In the following, the effect of the shear force is formulated.

\begin{figure*}
\centering\includegraphics[width=\textwidth]{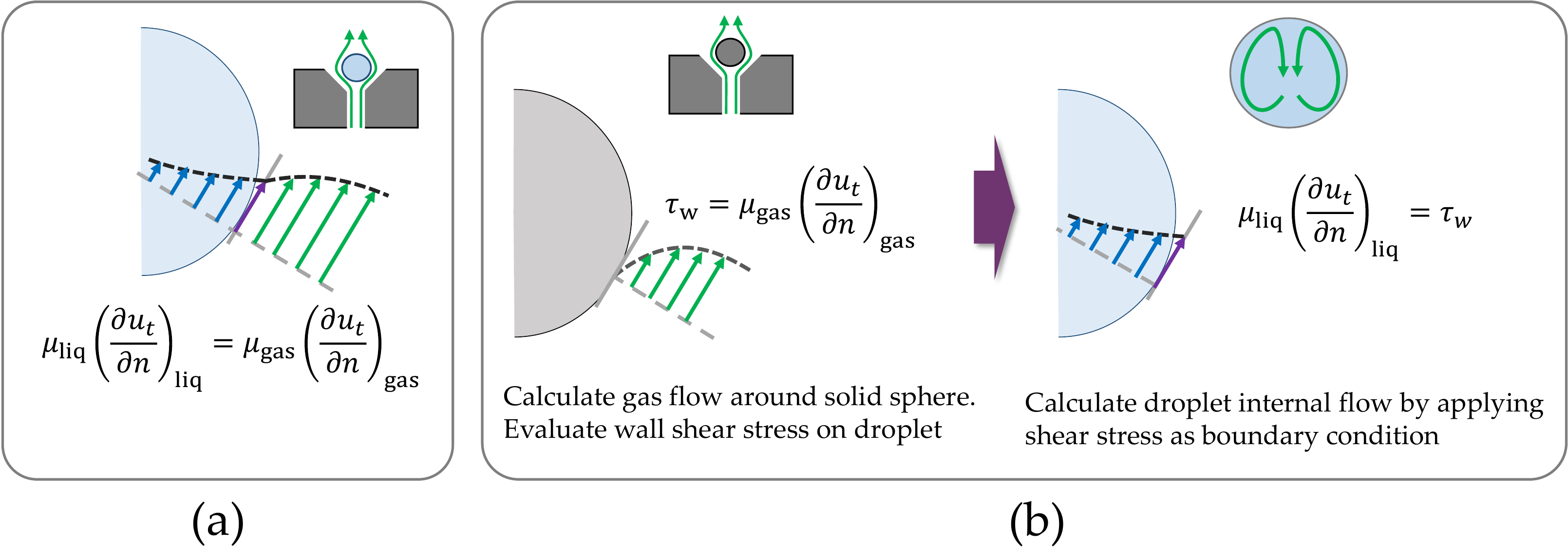}
\caption{\label{fig-ADL_shearBC} 
(a) Stress equilibrium on the liquid/gas interface.
(b) Approximated model in the present study.
}
\end{figure*}

At the liquid-gas interface,
the tangential stress balance can be written as
\begin{equation}
   \mu_\text{liq} \bol{D}_\text{liq} \cdot \bol{n}
 = \mu_\text{gas} \bol{D}_\text{gas} \cdot \bol{n},  \label{eq-shear1}
\end{equation}
where $\bol{D} = \nabla\bol{u} + \nabla\bol{u}^T$ is the strain rate tensor.
The subscripts `liq' and `gas' indicate the liquid and gas phases, respectively.
To implement \cref{eq-shear1},
the liquid and gas phases must be coupled in some way, which requires considerable effort.
In this study, 
\cref{eq-shear1} is divided into the following two equations and coupled in a one-way sense, as shown in \cref{fig-ADL_shearBC}.
\begin{subequations}
\label{eq-stressBC_ADL}
\begin{gather}
 \bol{\tau}_w = \mu_\text{gas} \bol{D}_\text{gas} \cdot \bol{n}, \\
 \mu_\text{liq} \bol{D}_\text{liq} \cdot \bol{n} = \bol{\tau}_w,
\end{gather}
\end{subequations}

\subsection{Specific model for ESL}
In the ESL system shown in \cref{fig-All_systems}(c),
an electrically conducting liquid droplet is placed in a uniform electrostatic field, which is generated by two electrodes.
The electric potential is constant everywhere inside the droplet,
thus no convection is driven by the electric origin \cite{Huo2004}.
The electric charge distribution is nonuniform along the free surface,
which results in a surface deformation of the droplet.
Under normal gravity, a high voltage of electrostatic field is required to levitate the metal droplet, 
thus, the surface deformation may exceed the magnitude that cannot be neglected.
Conversely, under a microgravity environment, 
the droplet can be assumed to be a sphere because the electrostatic field is only required  for the positioning of the droplet.

In this study, the model for the ESL system is targeted to Electrostatic Levitation Furnace (ELF) on board the ISS.
Therefore, the droplet is assumed to be spherical.
As the driving force for the convection, the Marangoni effect is considered.
Because the alignment of the lasers in the ELF is not axisymmetric,
the thermal and flow fields are considered in three dimensions.

\subsection{Nondimensionalization}

\begin{table*}
\caption{\label{tab-scales} Scales for nondimensionalization. 
The symbol $\nu = \mu / \rho$ is the kinematic viscosity.
}
\centering
\begin{tabularx}{\textwidth}{lccCL}
\toprule
                          & \multicolumn{2}{c}{Symbols}                \\
\cline{2-3}
Variable                  & Dimensional        & Nondimensional        & Scale                             & Remarks           \\
\midrule
Length                    & $\bol{x}$          & $\bol{X}$             & $d$                               & Droplet diameter  \\
Velocity                  & $\bol{u}$          & $\bol{U}$             & $u_0 = \nu/d$                     &                   \\
Time                      & $t$                & $\tau$                & $t_0 = d^2 /\nu$                  &                   \\
Pressure                  & $p$                & $P$                   & $p_0 = \rho \nu^2 / d^2$          &                   \\
Temperature               & $T$                & $\Theta$              & $T_\ast$                          & Melting point     \\
Magnetic flux density     & $\bol{B}$          & $\widehat{\bol{B}}$   & $B_0 = \sqrt{\mu /\sigma_e d^2}$  &                   \\
Electric current density  & $\bol{J}$          & $\widehat{\bol{J}}$   & $J_0 = B_0 / \mu_0 d$             &                   \\
\bottomrule
\end{tabularx}
\end{table*}

\begin{table*}
\sisetup{retain-zero-exponent=true}
\caption{\label{tab-nondimNumbers} 
Definition of nondimensional numbers and their ranges calculated in the present study. 
The symbol $\bigcirc$ shown in the columns of levitation methods indicates that the corresponding nondimensional number is involved in the model.
}
\centering
\begin{tabularx}{\textwidth}{lCccccCC}
\toprule
 & &  \multicolumn{3}{c}{Involved} &  & \multicolumn{2}{c}{Ranges} \\
\cline{3-5} \cline{7-8}
Name                      & Symbol              & EML         & ADL         & ESL        && Lower bound     & Upper bound     \\
\midrule                                                                              
Prandtl number            & $\Pr$               & $\bigcirc$  & $\bigcirc$  & $\bigcirc$ && \num{2.7e-2}  & \num{9.1e-2}  \\
Galilei number            & $\Ga$               & $\bigcirc$  & $\bigcirc$  & ---        && \num{3.4e+6}  & \num{7.7e+11} \\
Marangoni number          & $\Ma$               & $\bigcirc$  & $\bigcirc$  & $\bigcirc$ && \num{8.0e+3}  & \num{2.3e+5}  \\
Biot number               & $\Bi$               & $\bigcirc$  & $\bigcirc$  & $\bigcirc$ && \num{1.9e-4}  & \num{4.6e-2}  \\
Planck number             & $\Pl$               & $\bigcirc$  & $\bigcirc$  & $\bigcirc$ && \num{6.2e0}   & \num{1.3e+2}  \\
Laser power number        & $\La$               & ---         & $\bigcirc$  & $\bigcirc$ && \num{7.4e-3}  & \num{1.4e-1}  \\
Magnetic Prandtl number   & $\Pm$               & $\bigcirc$  & ---         & ---        && \num{4.0e-7}  & \num{8.8e-7}  \\
Eckert number             & $\Ec$               & $\bigcirc$  & ---         & ---        && \num{1.2e+14} & \num{8.0e+14} \\
Magnetic number           & $\Mg$               & $\bigcirc$  & ---         & ---        && \num{2.7e+2}  & \num{1.3e+3}  \\
Shielding parameter       & $\Sp$               & $\bigcirc$  & ---         & ---        && \num{3.2e+1}  & \num{2.8e+2}  \\
Jet Reynolds number       & $\Rey_\text{jet}$   & ---         & $\bigcirc$  & ---        && \num{1.5e+3}  & \num{6.5e+3}  \\
Viscosity ratio           & $\nu_\ast$          & ---         & $\bigcirc$  & ---        && \num{1.3e+1}  & \num{2.4e+1}  \\
Reynolds number           & $\Rey$              & $\bigcirc$  & $\bigcirc$  & $\bigcirc$ && \multicolumn{2}{c}{evaluated from results} \\ 
\bottomrule
\end{tabularx}
\end{table*}


All the governing equations and boundary conditions are
nondimensionalized using scales listed in \cref{tab-scales}.
Nondimensional governing equations and boundary conditions are written as follows.
\begin{subequations}
\begin{gather}
 \nablaHat\cdot \bol{U} = 0,      \label{eq-continuity_N} \\
 \pdif{\bol{U}}{\tau} + \nablaHat \cdot \left( \bol{U}\bol{U} \right) 
 = -\nablaHat P + \nablaHat^2 \bol{U} + \Ga\, \bol{e}_g + \widehat{\bol{f}}_m,  \label{eq-NS_N} \\
 \pdif{\Theta}{\tau} + \nablaHat \cdot \left( \bol{U} \Theta \right) = \frac{1}{\Pr} \nablaHat^2 \Theta + \widehat{q}_m,   \label{eq-energy_N}
\end{gather}
\end{subequations}
\begin{subequations}
\begin{gather}
\left( \nablaHat\bol{U} + \nablaHat\bol{U}^T \right) \cdot \bol{n} = \frac{\Ma}{\Pr} \left( \bol{I} - \bol{n}\bol{n} \right) \cdot \nablaHat \Theta,   \label{eq-nonDimMa} \\
-\nablaHat \Theta \cdot \bol{n} =  \Bi \left( \Theta - \Theta_a \right) + \frac{1}{\Pl} \left( \Theta^4 - \Theta_a^4 \right) + \La\, \widehat{W} \left( \bol{X} \right),  \label{eq-nonDimHeat}
\end{gather}
\end{subequations}
where the symbols with a hat $\widehat{\cdot}$ are
nondimensional versions of the operator or variables corresponding to the symbols without hats.
The terms for the Lorentz force and Joule heat are written as
\begin{subequations}
\label{eq-nonDim_EM}
\begin{gather}
\widehat{\bol{f}}_m = \frac{1}{\Pm} \widehat{\bol{J}} \times \widehat{\bol{B}}, \\
\widehat{q}_m       = \frac{1}{\Pm^2 \Ec} \left| \widehat{\bol{J}} \right|^2.
\end{gather}
\end{subequations}
The nondimensional form of the governing equation for magnetic potential $\widehat{\bol{A}}$ and its solution
are written as
\begin{gather}
\nablaHat^2 \widehat{\bol{A}} = -i\, \Sp\, \widehat{\bol{A}}, \\
\widehat{A}_{\varphi} \left(\widehat{R}, \theta \right) = 
 \frac{\Mg  \sin \theta_s }{2\sqrt{i\, \Sp\, \widehat{R} \widehat{R}_0}}  
 \sum_{n=1}^{\infty} C_n{I_{n+\frac{1}{2}}  \left(\widehat{R} \sqrt{i\, \Sp} \right)  P_n^{1}}  \left( \cos \theta \right).  \label{eq-solA2}
\end{gather}
All nondimensional numbers that appear in the model are defined as follows:
\begin{subequations}
\begin{alignat}{2}
&\text{Prandtl number}            &\,\,  \Pr             & = \frac{\nu }{ \alpha},                                             \\
&\text{Galilei number}            &\,\,  \Ga             & = \frac{g d^3 }{ \nu^3},                                            \\
&\text{Marangoni number}          &\,\,  \Ma             & = \frac{-\sigma_T T_\ast d }{ \mu \alpha},                          \\
&\text{Biot number}               &\,\,  \Bi             & = \frac{d h }{ \lambda},                                            \\
&\text{Planck number}             &\,\,  \Pl             & = \frac{\lambda }{ \sigma_\text{SB}\, \varepsilon\, T_\ast^3\, d},  \\
&\text{Laser power number}        &\,\,  \La             & = \frac{I_0 }{ \lambda T_\ast d},                                   \\
&\text{Magnetic Prandtl number}   &\,\,  \Pm             & = \sigma_e\, \mu_0\, \nu,                                           \\
&\text{Eckert number}             &\,\,  \Ec             & = \frac{d^2 c_p T_\ast }{ \nu^2},                                   \\
&\text{Magnetic number}           &\,\,  \Mg             & = \frac{I_s\, \mu_0 }{ B_0\, d},                                    \\
&\text{Shielding parameter}       &\,\,  \Sp             & = \omega\, \mu_0\, \sigma_e\, d^2,                                  \\
&\text{Jet Reynolds number}       &\,\,  \Rey_\text{jet} & = \frac{u_\text{jet}\, d }{ \nu_\text{gas}},                        \\
&\text{Viscosity ratio}           &\,\,  \nu_\ast        & = \frac{\nu_\text{gas} }{ \nu} ,                                    \\
&\text{Reynolds number}           &\,\,  \Rey            & = \frac{u_\text{max} d }{ \nu} .
\end{alignat}
\end{subequations}
The ranges for these nondimensional numbers in the present numerical simulation are summarized in \cref{tab-nondimNumbers}.
The lower and upper bounds are evaluated from the droplet sizes employed and the range of the physical properties listed in \cref{tab-properties}. 
Reynolds number $\Rey$ is defined based on the maximum velocity of the internal convection,
which is calculated from the numerical results.

\section{Numerical simulations}\label{sec-simulations}

\subsection{Implementation}
All numerical methods described in this manuscript are implemented on the open-source CFD toolbox
\href{https://www.openfoam.com}{OpenFOAM}.
In OpenFOAM, the basic conservative equations are implemented on standard ready-made solvers,
whereas some components of the model must be newly implemented.
In this study, 
the \texttt{buoyantPimpleFoam} is selected as a base solver for all the levitation systems.
The Marangoni effect \cref{eq-Marangoni} 
and the heat source by a laser \cref{eq-laserW} are implemented as modules of the boundary condition.
%

\subsection{Calculation conditions}

\begin{table*}
\sisetup{retain-zero-exponent=true}
\caption{\label{tab-properties} Materials and their thermo-physical properties considered in the present numerical simulations.}
\centering
\begin{tabularx}{\textwidth}{p{33mm}ccCCCC}
\toprule
                                  &                             &                                  &  \multicolumn{4}{c}{Material} \\
\cline{4-7}                       
                                  &                             &                                  &  Titanium alloy                       & Tungsten                            & Vanadium                             & Ruthenium                               \\ 
Property                          & Symbol                      & Unit                             &  \TiSixFour                           & \ce{W}                              & \ce{V}                               & \ce{Ru}                                 \\ 
\midrule
Melting point\cite{Ishikawa_2011} & $T_\ast$                    & \si{K}                           &  1923                                 & 3695                                & 2183                                 & 2607                                    \\
Density \cite{Ishikawa_2011}      & $\rho$                      & \si{kg/m^3}                      &  \num{4.150E+03}                      & \num{1.643E+04}                     & \num{5.460E+03}                      & \num{1.075E+04}                         \\
Viscosity \cite{Ishikawa_2011}    & $\mu$                       & \si{\pascal\second}              &  \num{2.38E-03}                       & \num{6.9E-03}                       & \num{4.3E-03}                        & \num{6.1E-03}                           \\
Kinematic  viscosity              & $\nu = \mu / \rho$          & \si{m^2/s}                       &  \num{5.73e-7}                        & \num{4.20e-7}                       & \num{7.88e-7}                        & \num{5.67e-7}                           \\
Specific heat                     & $c_p$                       & \si{\joule/\kilo\gram\kelvin} &  \num{5.230e2} \cite{Holfelder_2020}  & \num{2.88e2} \cite{Pottlacher_1999} & \num{8.431e2} \cite{Pottlacher2007}  & \num{3.552e2} \cite{Paradis2004}        \\
Thermal conductivity              & $\lambda$                   & \si{\watt/\meter\kelvin}      &  \num{1.88e1} \cite{Mohr_2020}        & \num{6.20e1} \cite{Pottlacher_1999} & \num{3.98e1}  \cite{Pottlacher2007}  & \num{7.96e2}  \cite{Ho1972}             \\
Thermal diffusivity               & $\alpha = \lambda/\rho c_p$ & \si{m^2/s}  &  \num{8.66e-6}                        & \num{1.31e-5}                       & \num{8.65e-6}                        & \num{2.08e-5}                           \\
Electrical conductivity           & $\sigma_e$                  & \si{S/m}    &  \num{5.620E5} \cite{Joshi_2017}      & \num{8.453e5} \cite{White1997}      & \num{7.402E5} \cite{Desai_1984}      & \num{1.236E6} \cite{Arblaster_2016}     \\
Prandtl number                    & $\Pr = \nu/\alpha$          & N.D.        &  \num{6.62e-2}                        & \num{3.21e-2}                       & \num{6.74e-2}                        & \num{1.21e-2}  \\
Emissivity                        & $\varepsilon$               & N.D.        &  {0.500} \cite{Milo_evi__2012}        & {0.360} \cite{Allen_1960}           & {0.332} \cite{Cezairliyan_1974}      & {0.320} \cite{Milo_evi__2015}           \\
Temperature coefficient of surface tension \cite{Ishikawa_2011}
& \multirow{2}{*}{$\sigma_T$} 
& \multirow{2}{*}{\si{N/m K}} 
& \multirow{2}{*}{\num{-1.90E-4}} 
& \multirow{2}{*}{\num{-3.10E-4}} 
& \multirow{2}{*}{\num{-2.70E-4}} 
& \multirow{2}{*}{\num{-2.40E-4}}    \\
\bottomrule
\end{tabularx}
\end{table*}


The implemented solvers for the three levitation systems were executed 
by changing the physical properties of materials, the size of droplets, and the driving conditions.
As target materials, four metals of 
titanium alloy \TiSixFour, tungsten W, vanadium V, and ruthenium Ru were selected.
Their physical properties are summarized in \cref{tab-properties}.
\TiSixFour was selected because it is a widely used material for metal additive manufacturing.
Other materials were selected focusing on the distinction in their physical properties:
a large density of tungsten,
a small viscosity of vanadium,
and large thermal diffusivity of ruthenium.
Concerning the droplet size, two values were selected.
For the EML case, the droplet diameter was selected as $d = \SI{6}{mm}$ and \SI{12}{mm}.
For the ADL case,
the droplet diameter was selected as $d = \SI{2}{mm}$ and \SI{3}{mm} for all materials,
and for the ESL case,
$d = \SI{1.5}{mm}$ and \SI{2}{mm} were selected.

The laser heating power $I_0$ was
determined so that the temperature reaches the melting point $T_\ast$,
based on the estimation of the temperature at the heat equilibrium.
At the heat equilibrium,
the total heat gains $Q_\text{in}$ and heat loss $Q_\text{out}$ must be balanced.
In the EML model,
the local heat gain is caused by Joule heat, and $Q_\text{in}$ can be evaluated by volume integration
\begin{equation}
Q_\text{in}^\text{Joule} = \int_V \frac{\left| \bol{J} \right|^2}{\sigma_e} dV,
\end{equation}
where $\bol{J}$ can be simply calculated by the analytical solution of electromagnetic field described in \cref{sec-EML_model}.
In the models of ADL and ESL,
the heat gain is caused by the laser heating applied as the boundary conditions.
Because all the laser power is assumed to be absorbed in the droplet,
the total heat gain is simply $Q_\text{in} = I_0$ in those cases.
The total heat loss $Q_\text{out}$ can be evaluated by surface integration of heat flux as
\begin{equation}
Q_\text{out} =  \int_S  \left[  h \left(T  - T_a \right) + \sigma_\text{SB} \epsilon \left( T^4 - T_a^4 \right) \right] dS. \label{eq-Qout}
\end{equation}
To analytically evaluate \cref{eq-Qout}, 
we assume that the surface temperature $T$ is uniform and the convective heat loss can be neglected.
Under such assumption,  \cref{eq-Qout} can be written as
\begin{equation}
Q_\text{out} = \sigma_\text{SB} \varepsilon \left( T^4 - T_a^4 \right)\,  4\pi \left( \frac{d}{2} \right)^2,  \label{eq-Qout2}
\end{equation}
which enables rough estimation of the temperature at the heat equilibrium $Q_\text{in} = Q_\text{out}$ as
\begin{equation}
T_\infty = \left( \frac{Q_\text{in}}{\varepsilon \sigma_\text{SB} 4\pi \left( \frac{d}{2} \right)^2} + T_a^4 \right)^{\frac{1}{4}}.  \label{eq-Tinf}
\end{equation}
The laser input power $I_0$ was determined so that
the estimated temperature $T_\infty$ in \cref{eq-Tinf} become grater than the melting point $T_\ast$.

\subsection{Simulations for EML}

\begin{figure*}
\centering\includegraphics[width=0.8\textwidth]{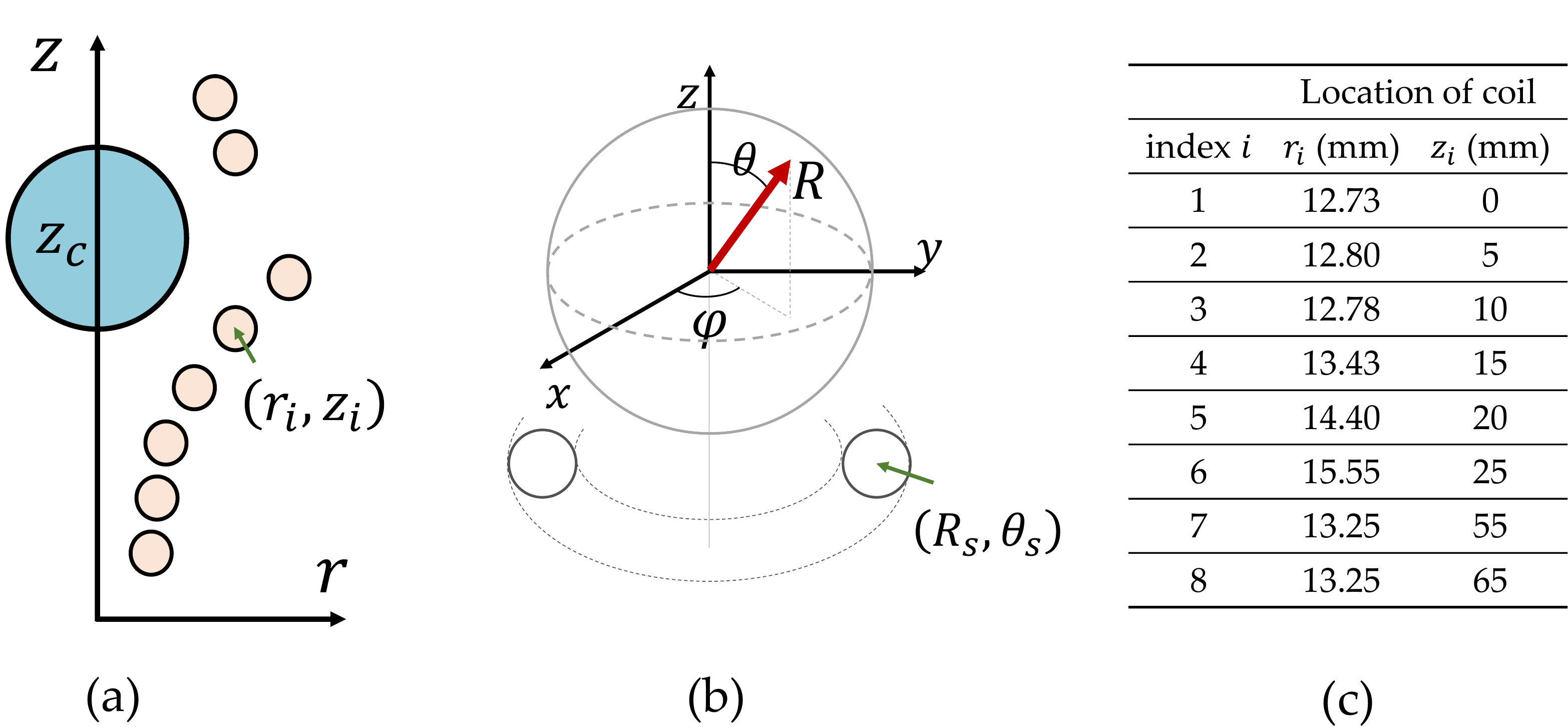}
\caption{\label{fig-coil} 
Configuration of the coil for the EML system.
(a) Schematics in axisymmetric cylindrical coordinates,
(b) coil location in spherical coordinates,
(c) detailed locations of the coils, 
which is determined according to the actual EML facility installed in the Chiba Institute of Technology (CIT) \cite{Ozawa_2021}.}
\end{figure*}

\subsubsection{Configuration and procedure}
For the EML system,
the calculations consisted of two steps.
First, the electromagnetic field was calculated by a separate solver,
then the Lorentz force $\bol{f}_m$ and Joule heat $q_m$ were obtained.
Using these $\bol{f}_m$ and $q_m$, the thermal and velocity field were calculated.
According to the experimental conditions in the Chiba Institute of Technology (CIT),
the AC frequency was selected as \SI{2e5}{Hz}, which corresponds to the angular frequency $\omega = \SI{1.257e6}{rad/s}$.
For the heat flux on the interface,
the convective heat transfer coefficient $h$ was selected based on the Nusselt number which is defined as
\begin{equation}
\Nu = \frac{h d}{\lambda_\text{gas}}. \label{eq-NuDef}
\end{equation}
The value of $\Nu$ has been predicted for several types of flow fields,
and for the laminar forced convection 
\begin{equation}
\text{Nu} \approx \num{0.664}\, \Rey_\text{gas}^{\frac{1}{2}}\, \PR_\text{gas}^{\frac{1}{3}}, \label{eq-NuPred}
\end{equation}
has been predicted and widely known \cite{Incropera2007}.
All the nondimensional numbers in \cref{eq-NuPred} are defined for the gas properties.
For the representative EML case in this study, these nondimensional numbers are evaluated as
$\PR_\text{gas} \approx \num{0.62}$ and
$\Rey_\text{gas} \approx \num{13} $,
thus the Nusselt number can be evaluated as
$\Nu \approx \num{2.0}$ which is corresponding to $h = \SI{10}{\watt\per\meter^2\kelvin}$.

In the calculation of the electromagnetic field for the EML system,
the detailed locations of coils$(R_s, \theta_s)$ in \cref{eq-solA} are required.
Although the actual coil is helically wounded, 
it is modeled by multiple axisymmetric filaments, as shown in \cref{fig-coil}.
The detailed locations of coils $(r_i, z_i)$ were determined
from the actual EML facility in the CIT \cite{Ozawa_2021}.
$z_c$ is the axial coordinate of the droplet center.
In real phenomena, the position of the droplet is 
determined from the balance between Lorentz force and the droplet weight under the applied electrical current amplitude $I_s$.
In the present numerical simulations, 
the force equilibrium was found using an iterative calculation by varying droplet position $z_c$ and/or amplitude $I_s$.
In the following sections, the results of force balances are described first,
then the droplet internal flows are shown.

\subsubsection{Force balances}
\begin{figure*}
\begin{tabular}{cc}
\includegraphics[width=0.55\textwidth]{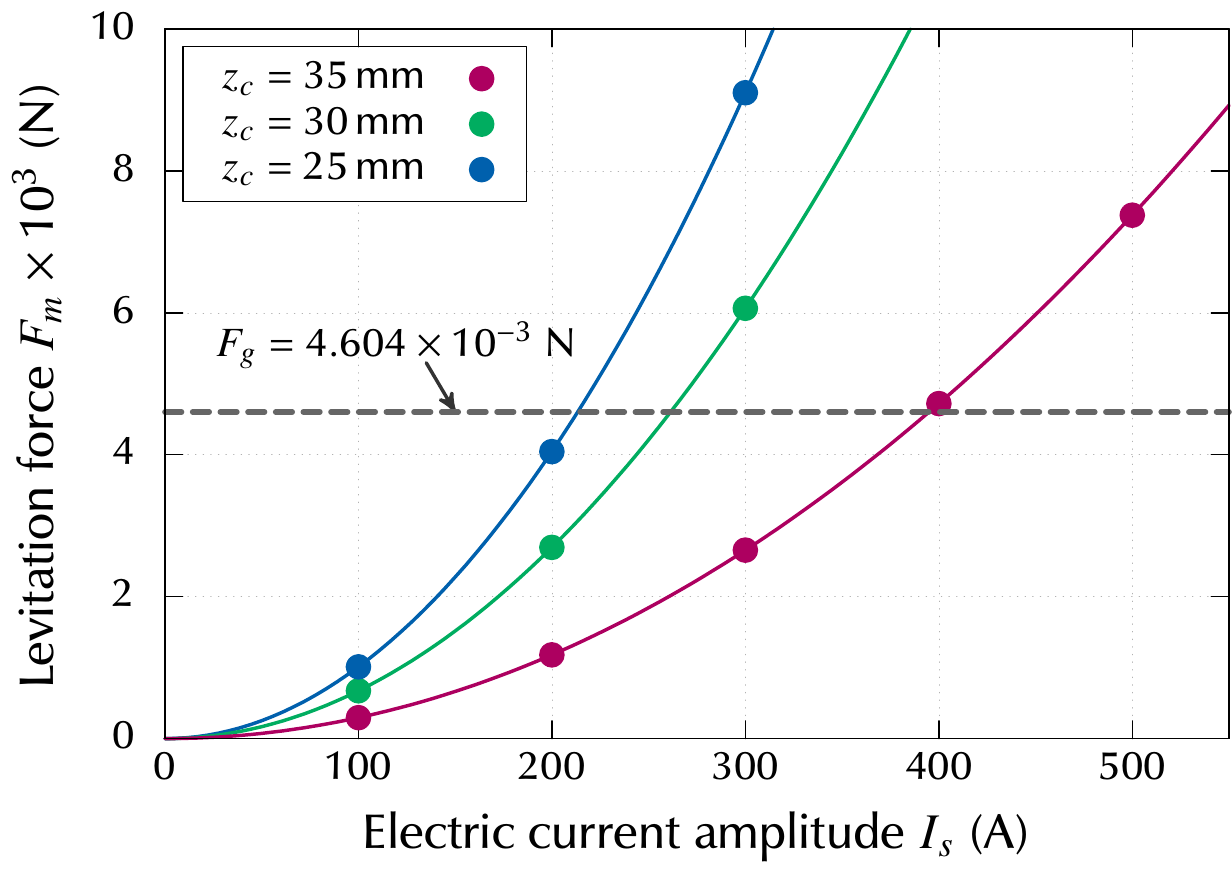} &
\includegraphics[width=0.418\textwidth]{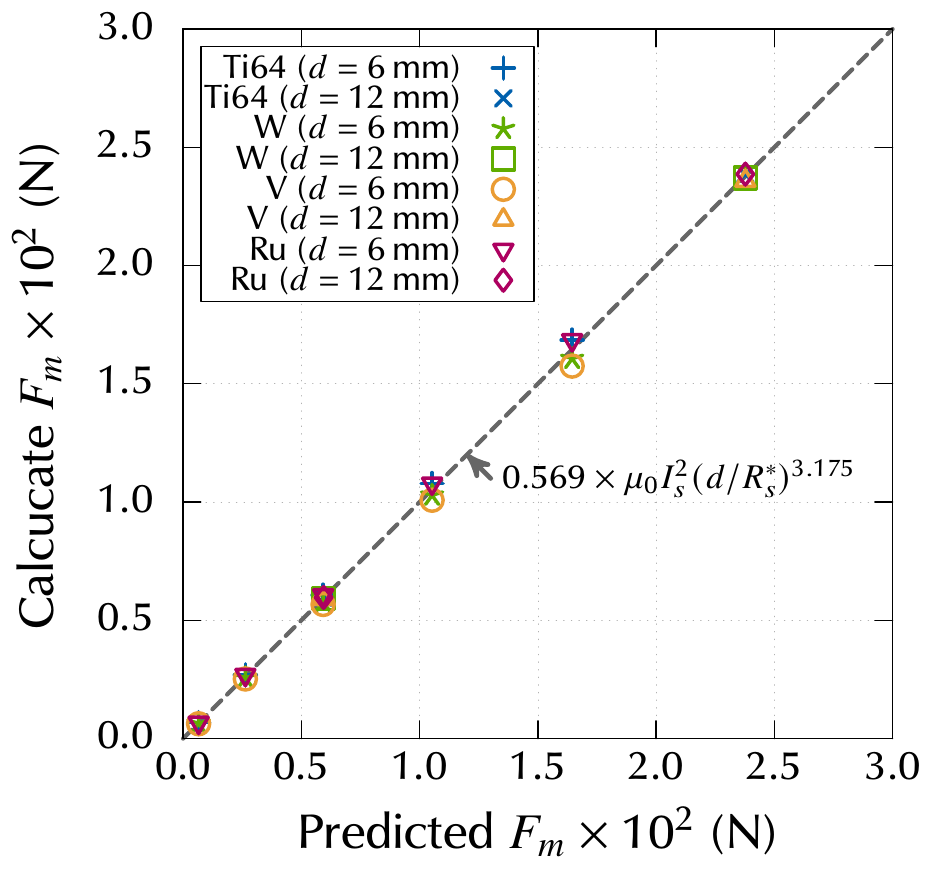} \\
(a) & (b)
\end{tabular}
\caption{\label{fig-EML_force} Levitation force $F_m$ as a function of electric current amplitude $I_s$.
(a) For the case of \TiSixFour of diameter $d=\SI{6}{mm}$. 
Circles represent the calculated results, and solid lines are fitted quadratic functions. 
The different colors correspond to droplet position $z_c$.
The gray dashed line represents droplet weight $F_g = m g$.
(b) Comparison between calculated and predicted levivation force $F_m$ for the case of $z_c = \SI{30}{mm}$.
The vertical axis represents $F_m$ calculated by the electro-magnetic model,
whereas the horizontal axis represents the values predicted by 
$0.569\times \mu_0 I_s^2 (d/R_s^\ast)^{3.175}$ (\cref{eq-Fm_EML}).
}
\end{figure*}
In the EML system,
the droplet weight must be balanced with the Lorentz force, 
which depends on the parameters related to the electromagnetic field,
such as electric current amplitude $I_s$ and electrical conductivity $\sigma_e$.
Because the droplet internal flow is driven by the Lorentz force,
the levitation condition must be preliminarily determined.
In this study, 
calculations of the electromagnetic field were conducted by changing electric current amplitude $I_s$,
while the axial location of the droplet was kept constant as $z_c = \SI{30}{mm}$. 
The levitation force was evaluated by integrating the local Lorentz force $f_m$ as
\begin{equation}
F_m = \int_V \bol{f}_m \cdot \bol{e}_z dV.
\end{equation}
\Cref{fig-EML_force}(a) shows
the result of $F_m$ as a function of $I_s$ for the \TiSixFour of $d = \SI{6}{mm}$.
Circles represent the calculated results, and solid lines are fitted quadratic functions. 
The difference in the colors corresponds to the droplet position $z_c$.
The gray dashed line represents droplet weight $F_g = m g$.
The intersection of two lines correspond to equilibrium $F_m = F_g$.
When the electric current amplitude $I_s$ is constant
the levitation force increases with decreasing droplet position $z_c$.
Through this relation, the droplet position is automatically determined in real phenomena.

Here, let us consider how the levitation force is scaled.
From the dimension analysis of the electromagnetic model described in \cref{sec-EML_model},
the levitation force can be scaled by $\mu_0 I_s^2$.
In the formulation of the magnetic potential \cref{eq-solA}, 
the Lorentz force is proportional to $R_0 / R_s$,
which is the ratio of the droplet size ($d = 2R_0$) over the coil position $R_s$.
In this study, the coil size was not changed even for the case of different droplet sizes;
thus, the ratio $R_0 / R_s$ increases with increasing $d$.
Based on the above-considered scaling,
the levitation force can be represented by
\begin{equation}
F_m^\ast = \num{0.569} \cdot {\mu_0 I_s^2}  \left(\frac{d}{R_s^\ast}\right)^{\num{3.175}}, \label{eq-Fm_EML}
\end{equation}
where the coefficient and exponent are determined by least squares fitting for the case of $z_c = \SI{30}{mm}$.
$R_s^\ast$ is the representative length of the coil, which is selected as \SI{12.73}{mm} in this study.
\Cref{fig-EML_force}(b) shows the comparison between calculated and predicted levitation force $F_m$
for all materials and droplet sizes considered in this study.
The vertical axis represents $F_m$ calculated by the electro-magnetic model,
whereas the horizontal axis represents the values predicted by \cref{eq-Fm_EML}.
It can be regarded that the prediction by \cref{eq-Fm_EML} shows good agreement with the calculated values.
In the calculation of droplet internal convection, which is described in the next subsection,
the droplet position is kept constant as $z_c = \SI{30}{mm}$
and $I_s$ is determined such that the levitation force balances the droplet weight.

\begin{figure*}
\begin{tabular}{cc}
\includegraphics[width=0.48\textwidth]{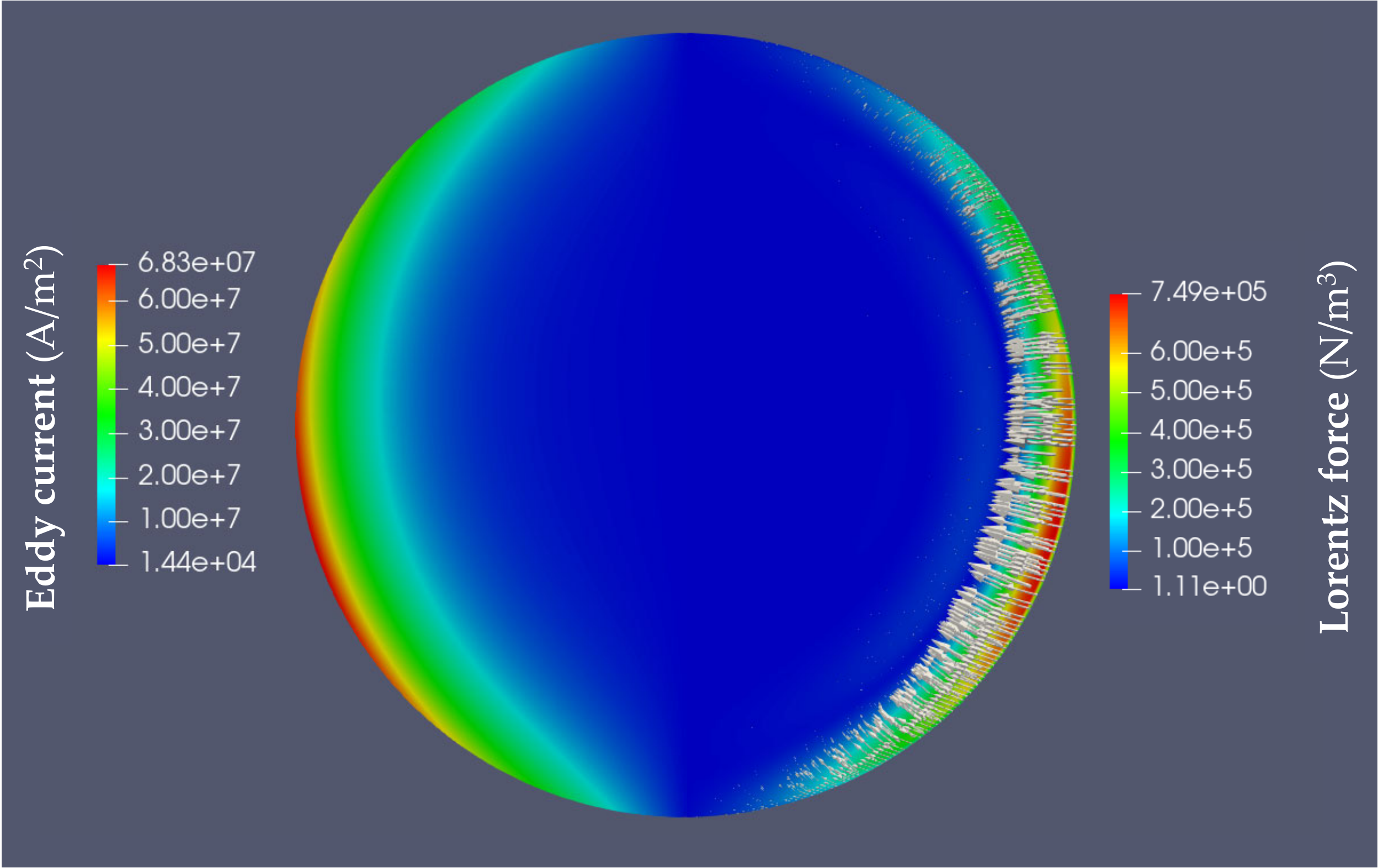} &
\includegraphics[width=0.48\textwidth]{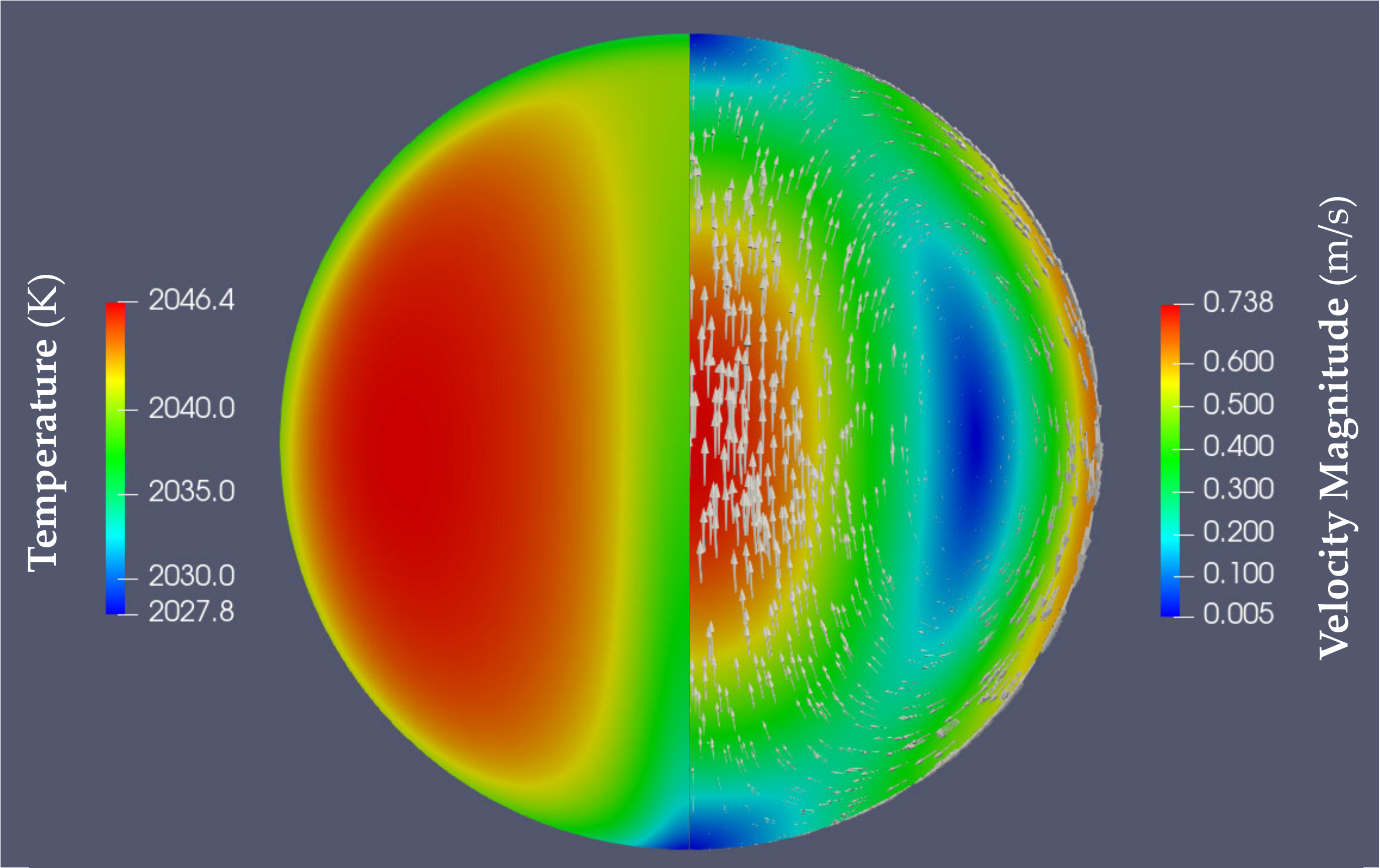} \\
(a) & (b)
\end{tabular}
\caption{\label{fig-EML_field}
The droplet internal flows for the EML system for \TiSixFour of $d=\SI{6}{mm}$.
The applied nondimensional numbers are
$\Pr = \num{6.6e-2}$,
$\Ga = \num{6.4e6}$,
$\Pm = \num{4.4e-7}$,
$\Ma = \num{1.1e5}$,
$\Ec = \num{1.2e14}$,
$\Bi = \num{3.2e-3}$,
$\Pl = \num{1.3e1}$,
$\Sp = \num{3.2e1}$,
$\Mg = \num{5.7e2}$.
(a) The eddy current $\bol{J}$ (left) and Lorentz force $\bol{f}_m$ (right), and
(b) temperature $T$ (left) and velocity magnitude (right).
The vectors are $\bol{f}_m$ and $\bol{u}$, respectively.
}
\end{figure*}

\subsubsection{Droplet internal convection}
Using the electromagnetic field for the equilibrium condition determined through the above-mentioned procedure,
the droplet internal convection is calculated.
\Cref{fig-EML_field} shows the calculated fields of
the eddy current $\bol{J}$, Lorentz force $\bol{f}_m$,
temperature $T$, and velocity $\bol{u}$.
The applied nondimensional numbers are
$\Pr = \num{6.6e-2}$,
$\Ga = \num{6.4e6}$,
$\Pm = \num{4.4e-7}$,
$\Ma = \num{1.1e5}$,
$\Ec = \num{1.2e14}$,
$\Bi = \num{3.2e-3}$,
$\Pl = \num{1.3e1}$,
$\Sp = \num{3.2e1}$,
$\Mg = \num{5.7e2}$.
The axisymmetic computaital domain is discretized by \num{21600} of meshes.
The calculation is conducted as time-dependent, 
and  \cref{fig-EML_field} is for the time after the fields are sufficiently developed.
In \cref{fig-EML_field}(a), the region of the strong Lorentz force $\bol{f}_m$ is concentrated near the lower part outside the droplet.
$\bol{f}_m$ is directed in the radially inward and axial upper sides,
and it drives the flow in this direction.
The eddy current $\bol{J}$ is also concentrated near the surface;
thus, the heat generated by the Joule heat $q =\left|\bol{J}\right|^2   / \sigma_e$ is localized there.
In \cref{fig-EML_field}(b),
the temperature field is averaged by the flow for a wide region.
The maximum velocity is $u_\text{max} = \SI{0.738}{m/s}$,
and the Reynolds number based on $u_\text{max}$ is evaluated as $\Rey = \num{7730}$.
Similar calculations are conducted for other materials and droplet sizes,
and the Reynolds number is an order of magnitude $\Rey \sim 10^4$,
as summarized in \cref{tab-results}.

\subsection{Simulations for ADL}

\subsubsection{Configuration and procedure}
\begin{figure*}
\centering\includegraphics[width=\textwidth]{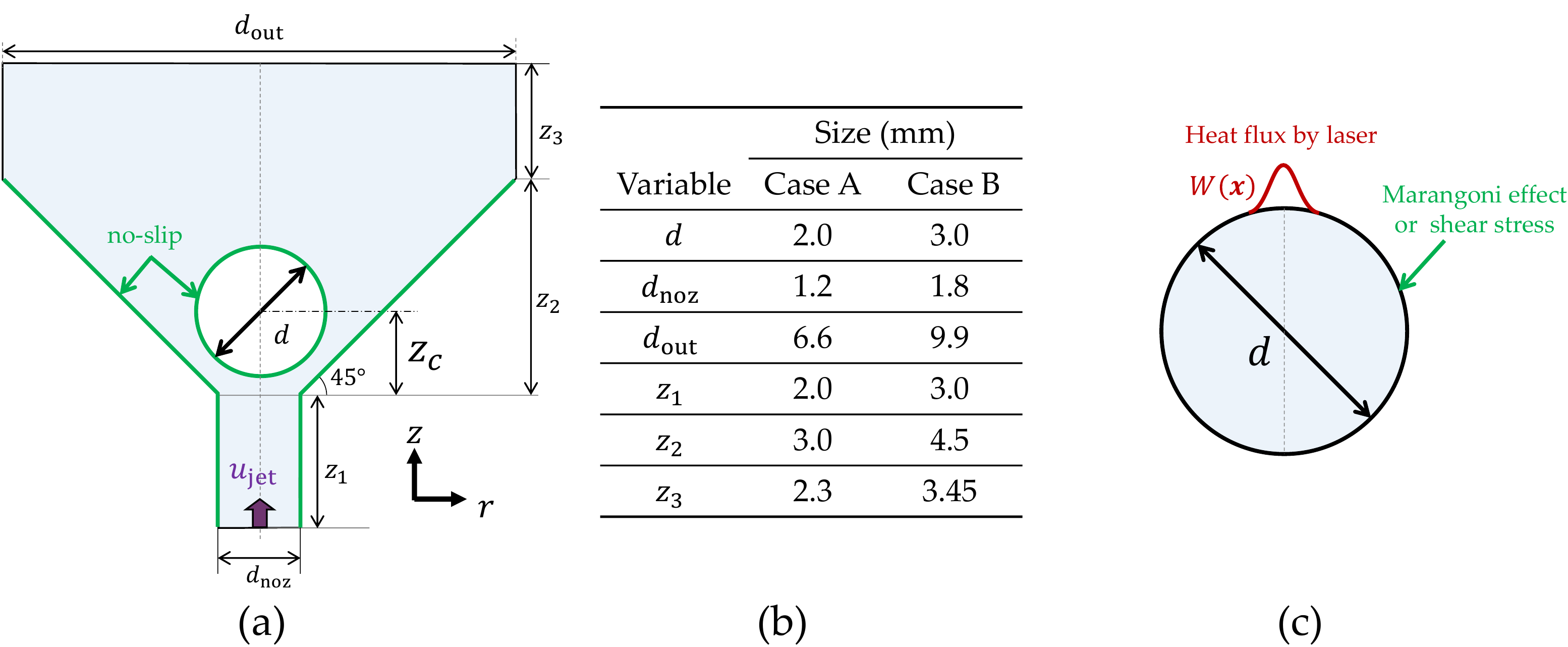}
\caption{\label{fig-ADL_domain} Computational domain for the ADL system.
(a) Gas flow domain,
(b) detailed sizes for the gas flow domain, and
(c) droplet domain.}
\end{figure*}

For the ADL system,
two types of convection driving forces, shear-induced and Marangoni convection were separately considered
as explained in \cref{sec-model_ADL}.
The shear-induced flow was calculated in two steps.
First, only the gas flow was considered by assuming the droplet surface as a no-slip rigid wall.
In the calculation of gas flow, a detailed spatial domain was designed as shown in \cref{fig-ADL_domain}(a,b), 
which was determined from the actual ADL facility installed in JAXA.
For the calculations of different droplet sizes $d$,
the diameter of the gas-jet nozzle $d_\text{noz}$ was also varied
while keeping the ratio constant as $d / d_\text{noz} =  5/3 $.
The axial location of the droplet center $z_c$ was considered as an adjustable parameter.
Similar to the EML system,
the position of the droplet was determined by the balance between the droplet weight and drag force from the gas jet,
which depend on the position of the droplet $z_c$ and flow rate of the gas jet.
In this study, 
the force equilibrium was found through an iterative calculation by changing droplet position $z_c$ and/or the volumetric flow rate $\phi_\text{jet}$.
After the equilibrium was found,
the wall shear stress $\tau_w$ on the droplet was evaluated,
then it was applied to the surface boundary condition in the calculation of droplet internal flows.

\begin{figure*}
\centering\includegraphics[width=0.6\textwidth]{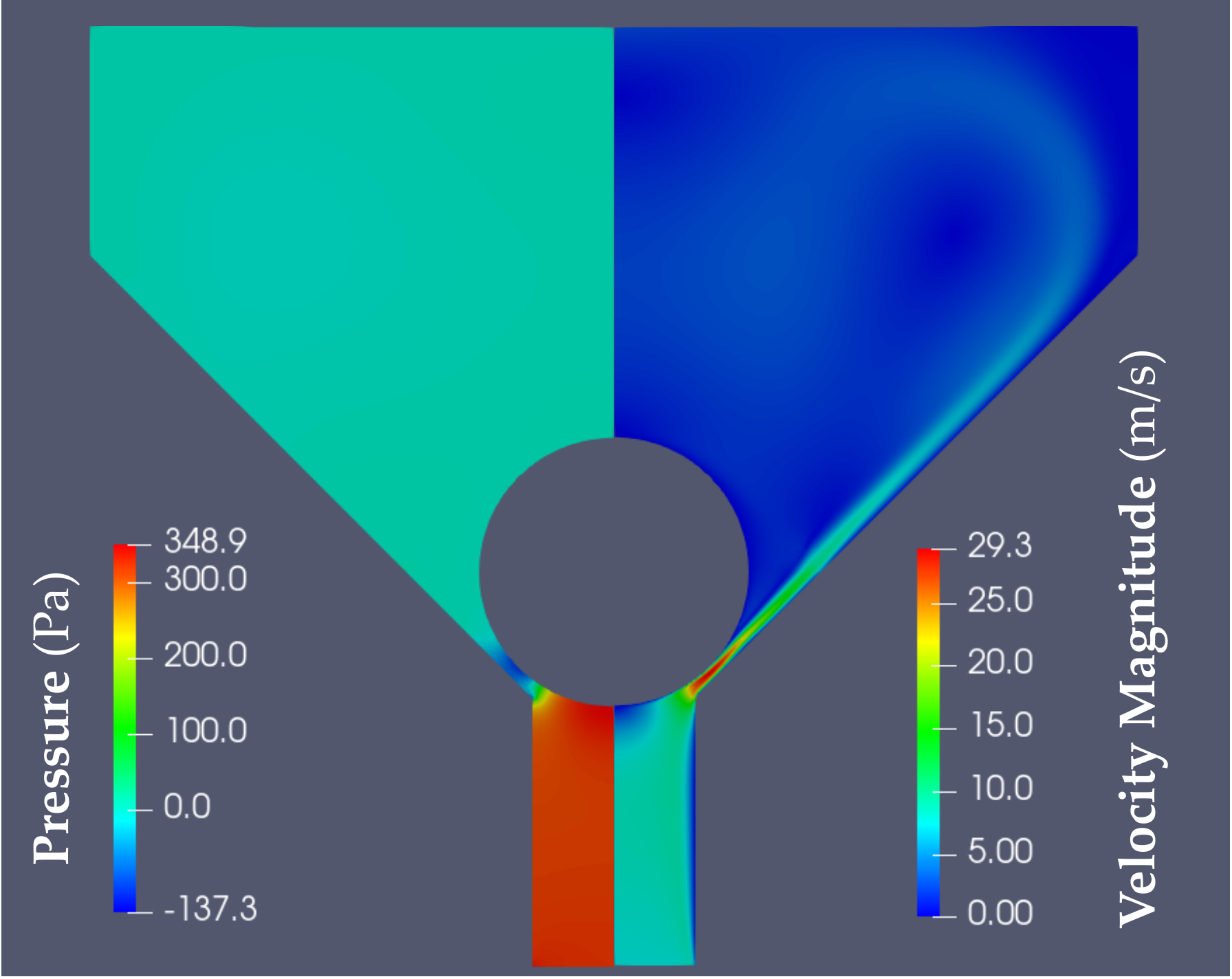}
\caption{\label{fig-ADL_gasFlow} 
Representative gas flow field for the droplet diameter of $d=\SI{2}{mm}$,
$z_c = \SI{0.94}{mm}$ and $\phi_\text{jet} = \SI{0.508}{L/min}$.
The corresponding jet velocity is $u_\text{jet} = \SI{7.49}{m/s}$.
Pressure (left) and velocity magnitude (right) fields .}
\end{figure*}

For the Marangoni convection,
only the droplet internal flow was calculated
by applying the heat flux and the Marangoni effect on the boundary condition.
In laser heating, only the upper surface was heated by a single laser.
The power of laser $I_0$ was determined such that the minimum temperature in the droplet
became larger than the melting point of the material.
For the heat flux on the interface,
the convective heat transfer coefficient $h$ was selected based on the Nusselt number \cref{eq-NuDef,eq-NuPred}
for the laminar forced convection, which is considered in the case of EML.
For the representative case in this study, these nondimensional numbers are evaluated as
$\PR_\text{gas} \approx \num{0.62}$ and
$\Rey_\text{jet} \approx \num{650} $,
thus the Nusselt number can be evaluated as
$\Nu \approx \num{14.4}$ which is corresponding to $h = \SI{430}{\watt\per\meter^2\kelvin}$.
Although $h$ is dependent on $\Rey_\text{jet}$, which is not constant,
the value of $h$ is kept constant in all ADL simulations.

\subsubsection{Gas flow and levitation force}

\begin{figure*}
\centering
\begin{tabular}{cc}
\includegraphics[width=0.48\textwidth]{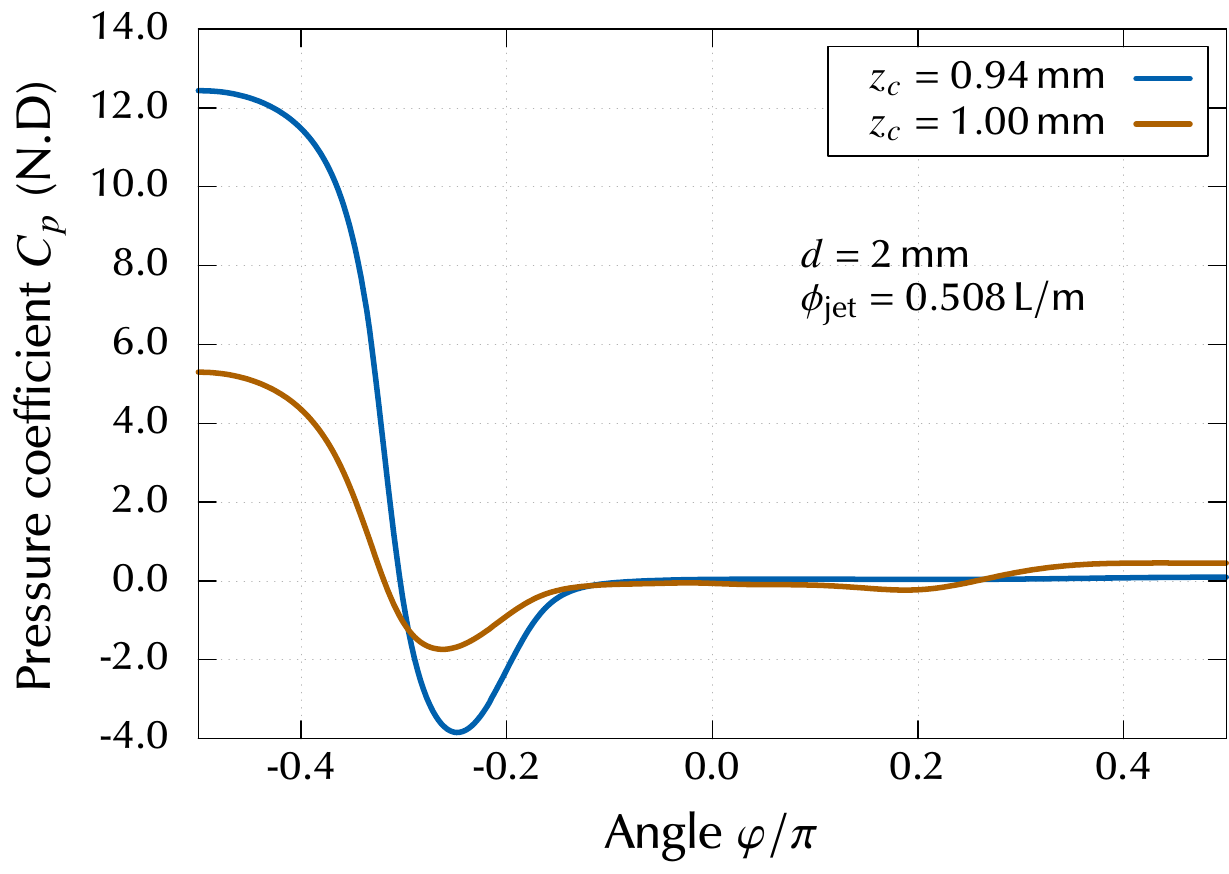} &
\includegraphics[width=0.48\textwidth]{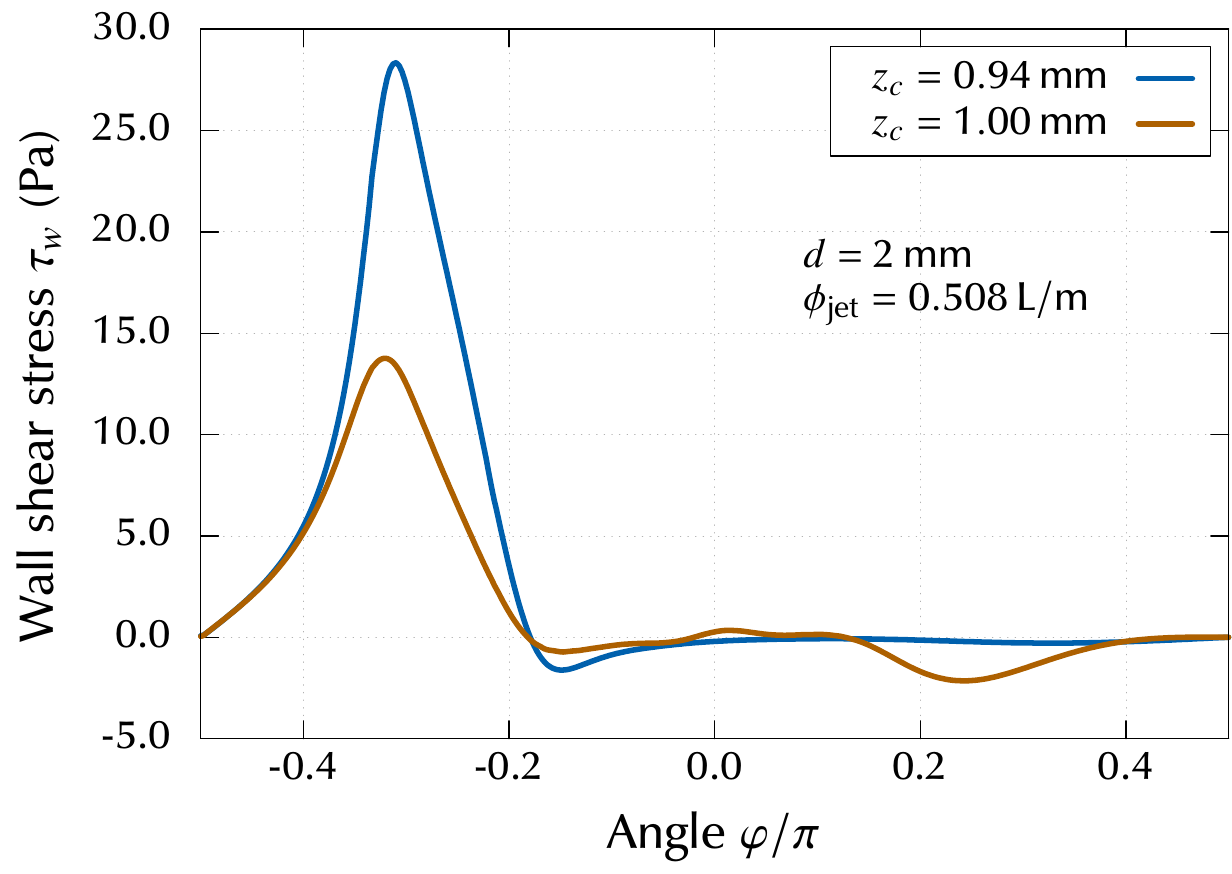} \\
(a) & (b)
\end{tabular}
\caption{\label{fig-ADL_P_WSS}
Pressure (a) and shear stress (b) distributions along the droplet wall
for the case of $d=\SI{2}{mm}$ and $\phi_\text{jet} = \SI{0.508}{L/min}$.
Two different droplet positions $z_c = \SI{0.94}{mm}$ and  $z_c = \SI{1.00}{mm}$ are compared.}
\centering
\begin{tabular}{cc}
\includegraphics[width=0.48\textwidth]{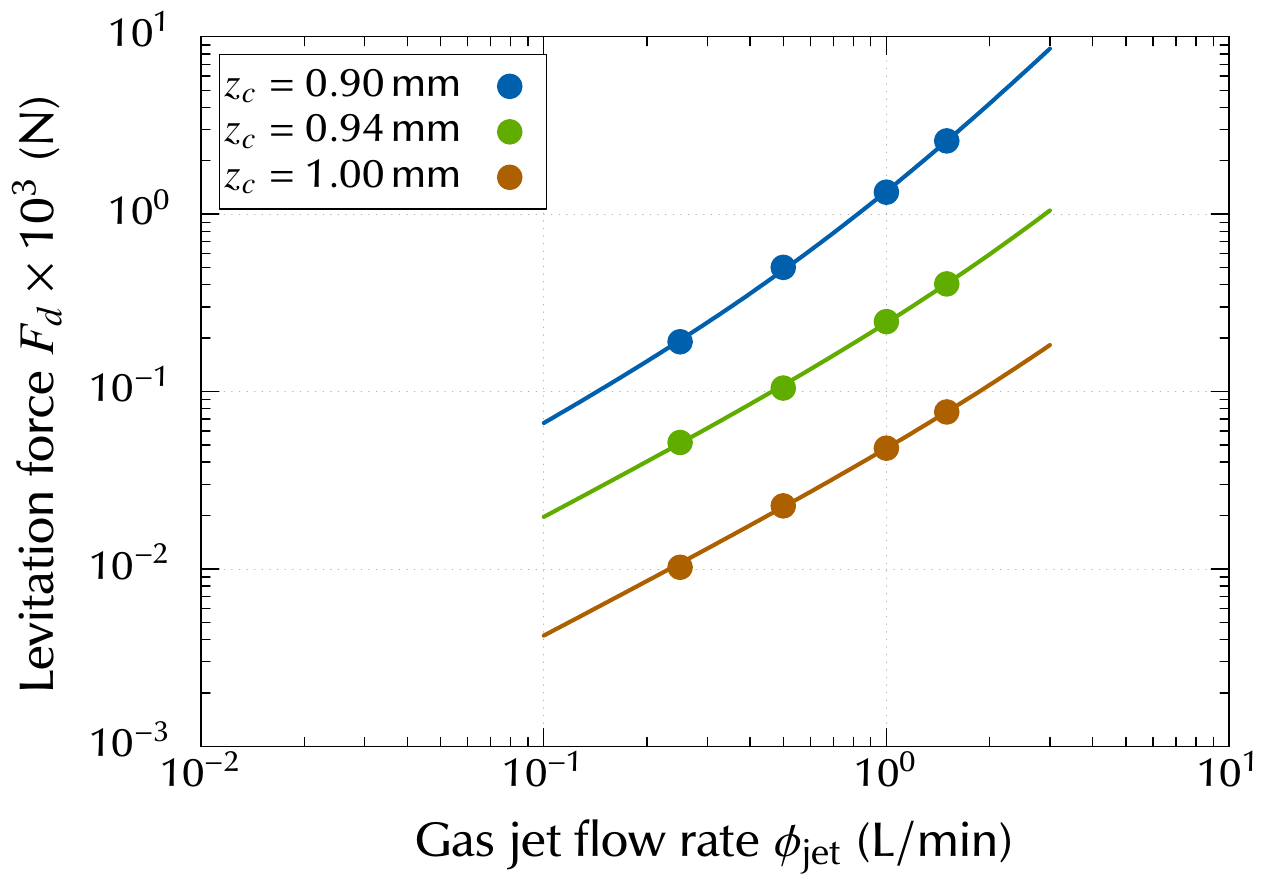} &
\includegraphics[width=0.48\textwidth]{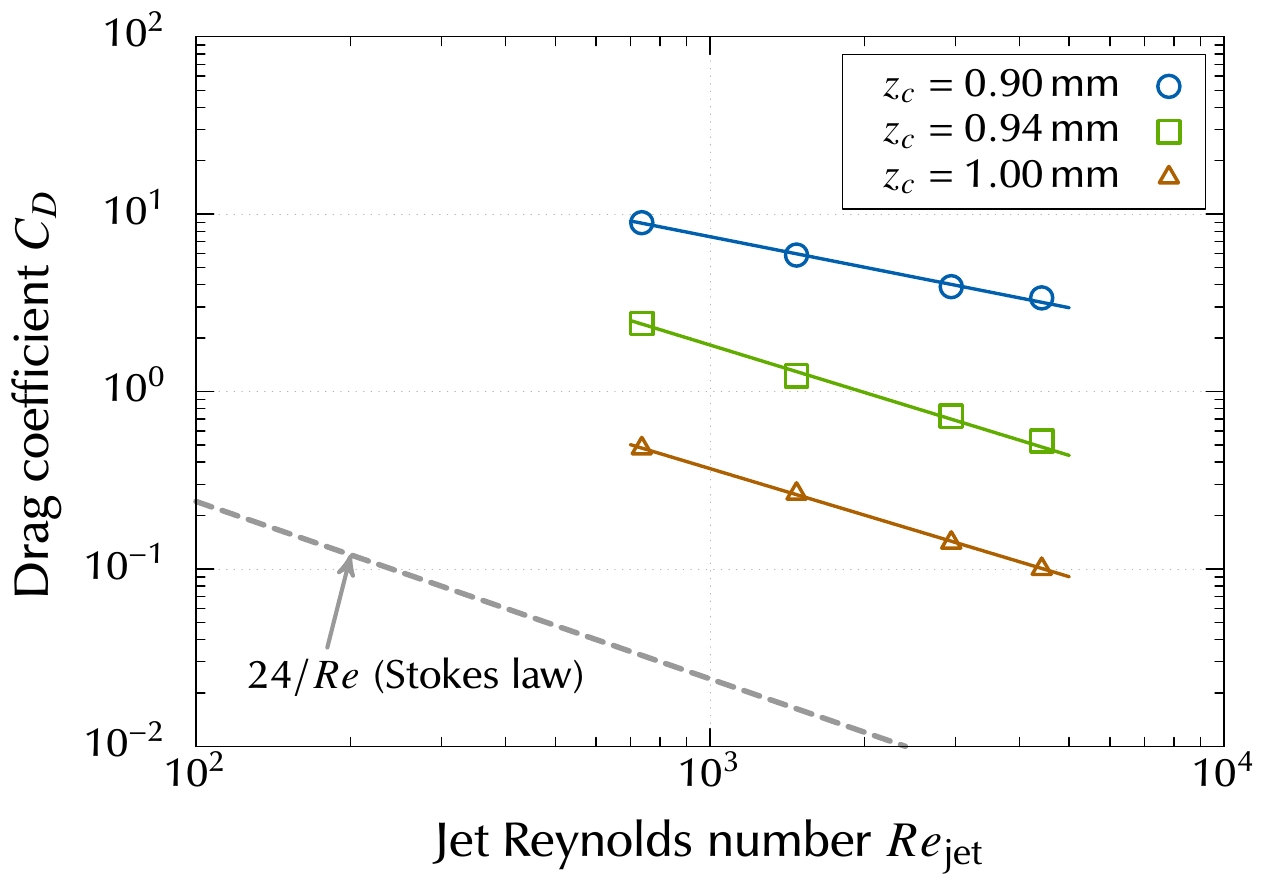} \\
(a) & (b) 
\end{tabular}
\caption{\label{fig-ADL_force} 
(a) Drag (levitation) force as a function of the gas-jet flow rate 
for the droplet size of $d=\SI{2}{mm}$.
The nondimensional droplet position is kept constant as $\widehat{z}_c = z_c / d = \num{0.47}$.
The circles indicate the calculated results, and solid lines are fitted quadratic functions.
(b) The drag coefficient $C_D$ as a function of the jet Reynolds number $\Rey_\text{jet}$. 
The solid lines indicate the fitted function
and the gray dashed line represents Stokes law for the drag on the sphere.
}
\end{figure*}

\Cref{fig-ADL_gasFlow} shows a representative gas flow field for the case of droplet diameter $d=\SI{2}{mm}$.
The droplet position is $z_c = \SI{0.94}{mm}$ and 
the volumetric flow rate is $\phi_\text{jet} = \SI{0.508}{L/min}$,
which corresponds to jet velocity $u_\text{jet} = \SI{7.49}{m/s}$.
The left and right contours indicate the pressure, velocity magnitude, respectively.
At the narrow gap between the droplet and nozzle, 
the velocity must be large due to continuity, thus the gauge pressure decreases.

The pressure distribution along the droplet wall is shown in \cref{fig-ADL_P_WSS}(a)
for the case of $d=\SI{2}{mm}$ and $\phi_\text{jet} = \SI{0.508}{L/min}$.
In the figure, the pressure is indicated by the pressure coefficient, which is defined as $C_p = p / p_d$, 
where $p_d = (\rho u_\text{jet}^2 /2)$ is the upstream dynamic pressure.
Two different droplet positions $z_c = \SI{0.94}{mm}$ and $z_c = \SI{1.00}{mm}$ are plotted for comparison.
The pressure distributions in two cases are significantly different.
The maximum pressure for the case of $z_c = \SI{0.94}{mm}$ is $p = \SI{348.9}{Pa}$,
whereas for the case of $z_c = \SI{1.00}{mm}$, $p = \SI{143.6}{Pa}$.
These values of stagnation pressure are much larger than the dynamic pressure at the inlet
$p_d = \SI{28.1}{Pa}$.
In addition, the high-pressure region is widely spread along the bottom of the droplet.
This strong pressure causes a large drag force.
This tendency becomes strong when the droplet is placed at a lower position.

To find the force equilibrium,
a series of gas flow calculations is executed
by changing $z_c$ and $\phi_\text{jet}$.
\Cref{fig-ADL_force}(a) shows drag force $F_d$ as a function of flow rate $\phi_\text{jet}$
for three different droplet positions $z_c$ by keeping the droplet diameter $d = \SI{2}{mm}$.
The circles represent the calculated results and solid lines are fitted quadratic functions.
By finding the intersection of the fitted curve and droplet weight,
the force equilibrium condition of $z_c$ and $\phi_\text{jet}$ can be determined.
For a constant jet flow rate $\phi_\text{jet}$,
the levitation force increases with decreasing droplet position $z_c$.

The drag force can be expressed by the relation between drag coefficient $C_D$ and the Reynolds number. 
Once $C_D$ is known, drag force $F_D$ can be evaluated for any conditions of droplet sizes $d$ and flow rates $\phi_\text{jet}$.
To this end, all the results of gas flow calculations are summarized 
using the drag coefficient and gas-jet Reynolds number defined as follows:
\begin{align}
\Rey_\text{jet} &= \frac{u_\text{jet} d}{\nu_\text{gas}}, \\
C_D            &= \frac{F_D}{p_d S_d}  , \\
p_d            &= \frac{1}{2} \rho_\text{gas} u_\text{jet}^2 , \\
S_d            &= \pi \left(\frac{d}{2}\right)^2,
\end{align}
where $p_d$ is a dynamic pressure and $S_d$ is the area of the nozzle outlet.
\Cref{fig-ADL_force}(b) shows the drag coefficient $C_D$ as a function of the jet Reynolds number $\Rey_\text{jet}$. 
The solid lines indicate the fitted function
and the gray dashed line is the Stokes law for the drag on the sphere.
In the \cref{fig-ADL_force}(b), 
the dependence of $C_D$ on $\Rey_\text{jet}$ is similar to that of Stokes law,
whereas the absolute value is highly dependent on the droplet position.

After the force balance was found,
the shear stress $\tau_w$ along the droplet wall was evaluated.
\Cref{fig-ADL_P_WSS}(b)
shows the shear stress $\tau_w$ along the droplet wall 
for the case $d = \SI{2}{mm}$ and $\phi_\text{jet} = \SI{0.508}{L/min}$
with the comparison of two droplet positions $z_c = \SI{0.94}{mm}$ and \SI{1.00}{mm}.
The two distributions of $\tau$ have similarities except for the absolute value.
The angles $\varphi$ where the $\tau_w$ take extrema are approximately the same for two cases of $z_c$.
The maximum value of $\tau_w$ is proportional to the stagnation pressure.
Using these distributions of $\tau_w$, the droplet internal convection is calculated.

In the calculation of droplet internal convection, which is discussed in the next subsection,
the droplet position is kept constant as $z_c = \SI{0.94}{mm}$
and the flow rate $\phi_\text{jet}$ is determined such that the levitation force balances with the droplet weight.

\subsubsection{Droplet internal convection}
\begin{figure*}
\centering
\begin{tabular}{cc}
\includegraphics[width=0.635\textwidth]{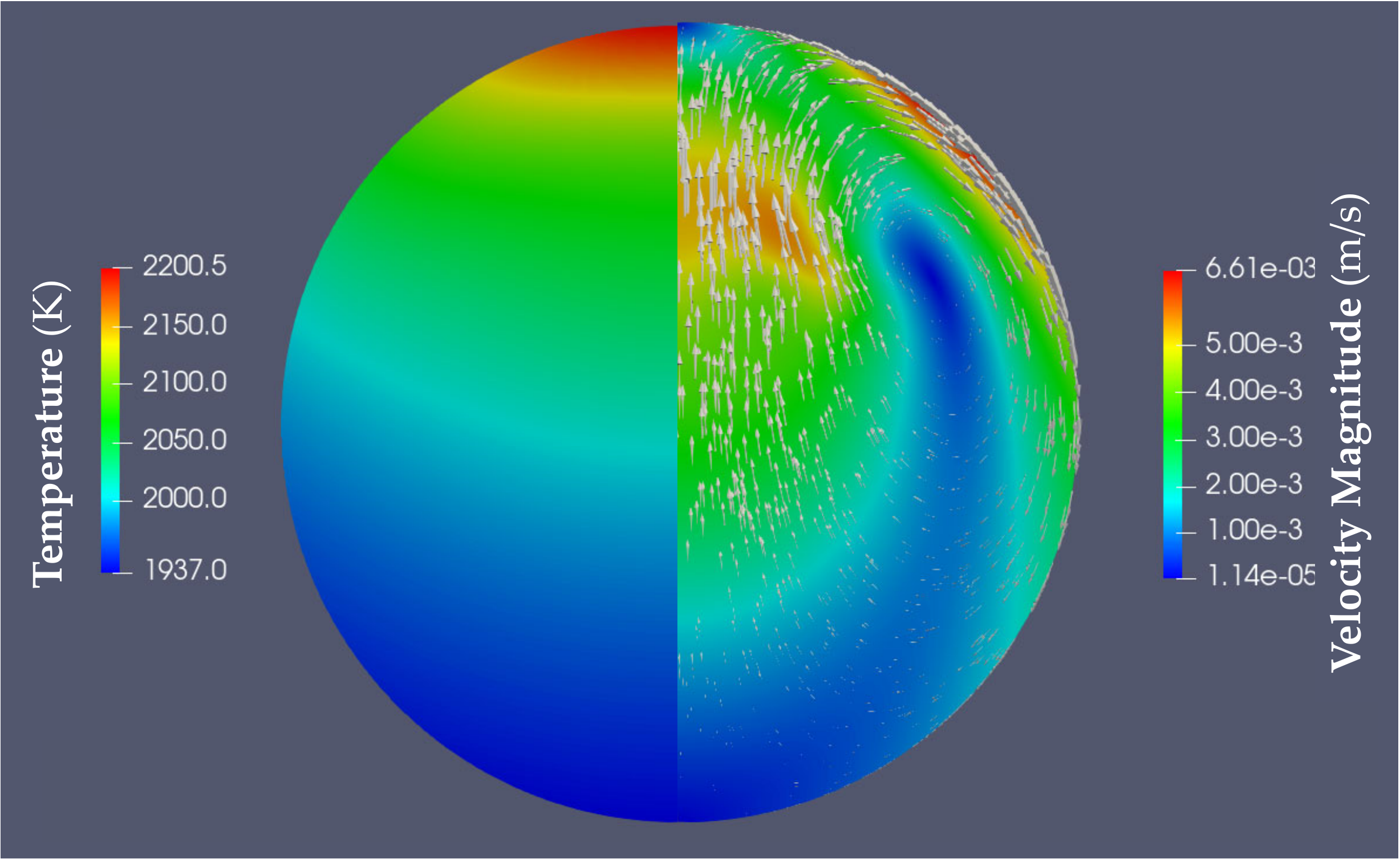} &
\includegraphics[width=0.325\textwidth]{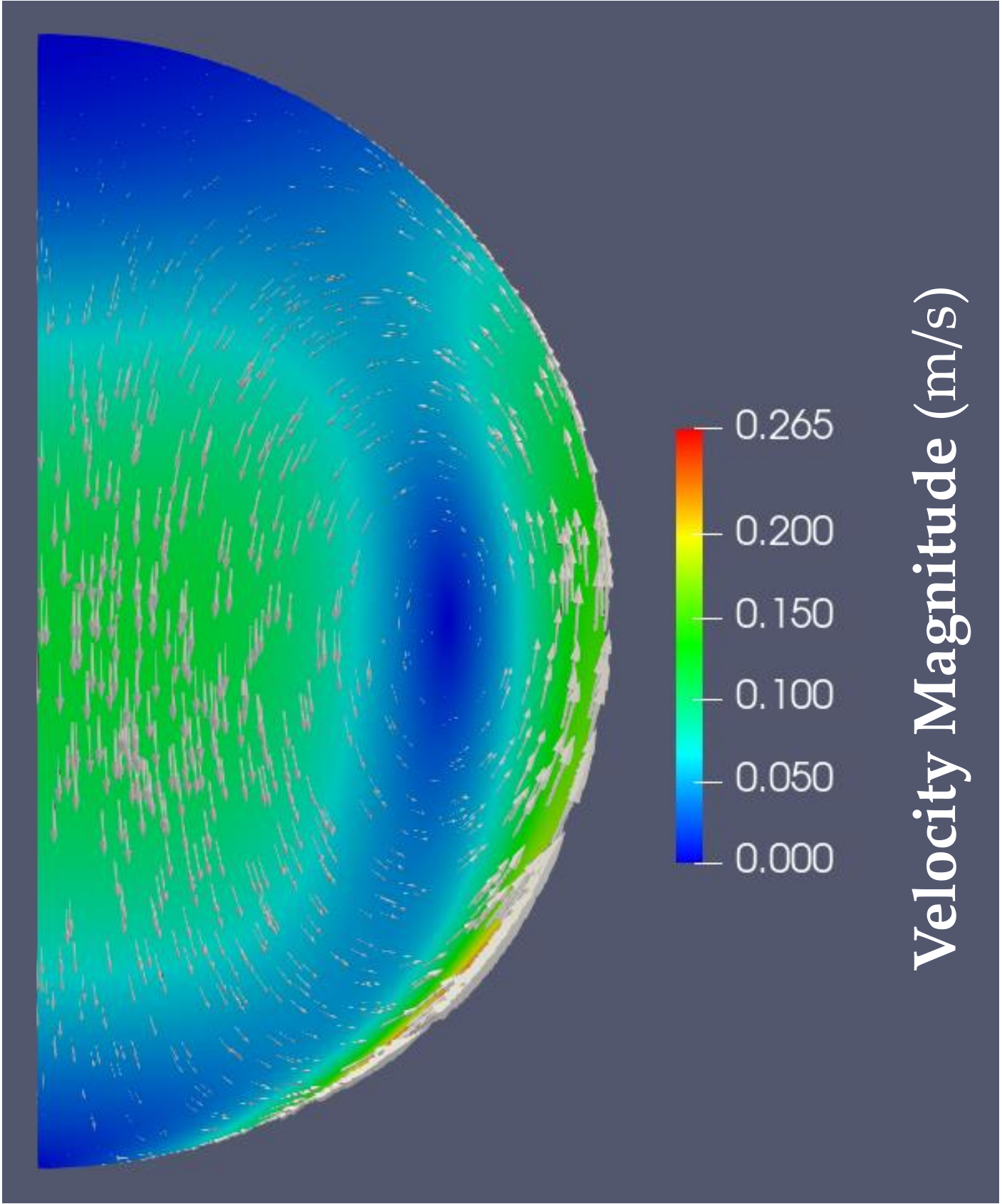} \\
(a) Marangoni convection & (b) Shear-induced convection
\end{tabular}
\caption{\label{fig-ADL_convection} 
Droplet internal convection for the ADL system
in the case of \TiSixFour with $d=\SI{2}{mm}$.
(a) Convection driven by the Marangoni effect due to laser heating.
The applied nondimensional numbers are
$\La = \num{1.4e-1}$,
$\Ma = \num{4.1e4}$,
$\Bi = \num{4.6e-2}$, and
$\Pl = \num{3.1e1}$.
The color contours on the left and right indicate temperature $T$
and velocity magnitude, respectively.
(b) Convection driven by the surface shear stress $\tau_w$ 
for the case of
$\Rey_\text{jet} = \num{1.5e3}$ and
$\nu_\ast = \num{1.7e1}$.
The color contour indicates the velocity magnitude.
}
\end{figure*}

\begin{figure*}
\centering\includegraphics[width=0.8\textwidth]{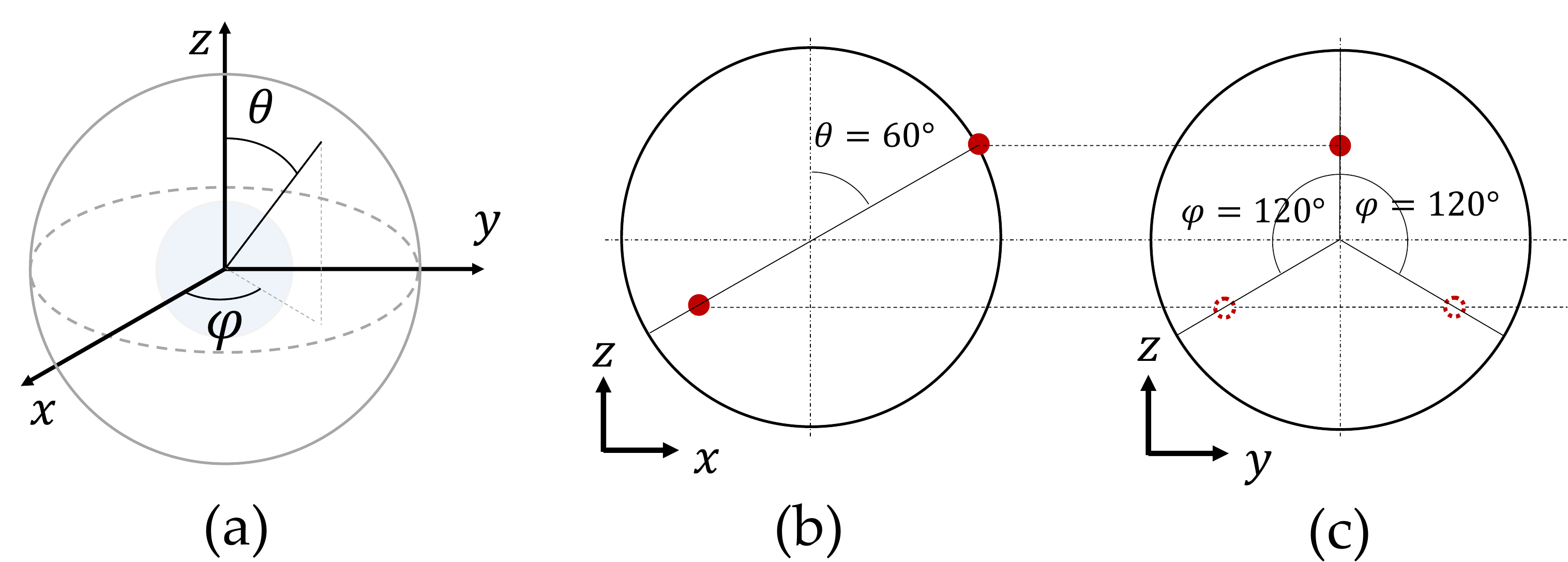}
\caption{\label{fig-ESL_laser} Configuration of the heating lasers for the ESL system.}
\end{figure*}

\begin{figure*}
\centering\includegraphics[width=0.7\textwidth]{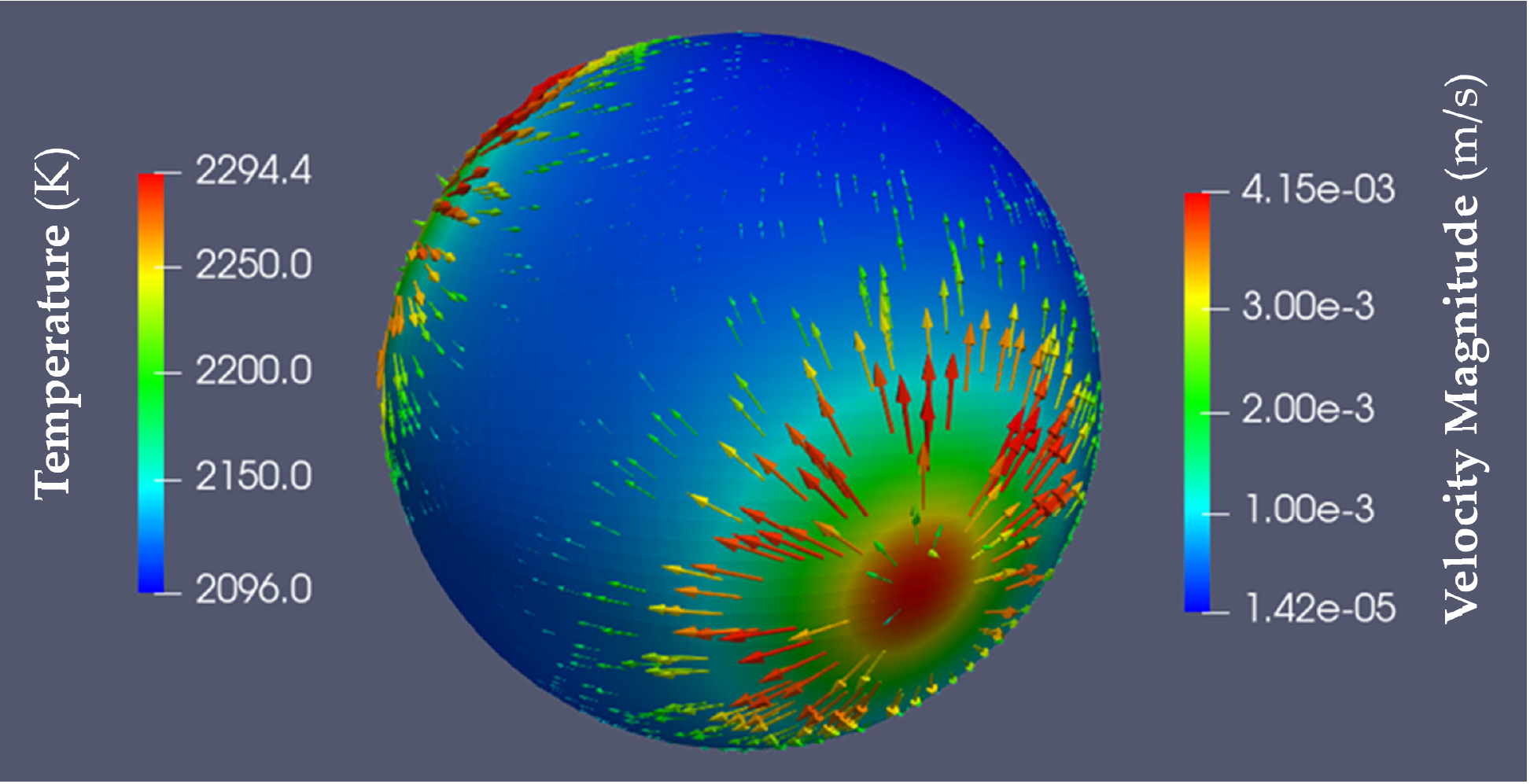}
\caption{\label{fig-ESL_field} 
Temperature (contour) and velocity (vectors) fields
of the droplet internal flow in the  ESL system
for the case of \TiSixFour with $d = \SI{2}{mm}$.
The applied nondimensional numbers are 
$\Pr = \num{6.6e-2}$,
$\Pl = \num{2.7e1}$,
$\Bi = \num{1.1e-3}$,
$\Ma = \num{4.2e4}$, and
$\La = \num{4.6e-2}$.
}
\end{figure*}


\begin{table*}
\small
\caption{\label{tab-results} Summary of simulation results. }
\begin{tabularx}{\textwidth}{cccCCCCCCCC}
\toprule
         &                & Units    & \multicolumn{2}{c}{\TiSixFour}     & \multicolumn{2}{c}{\ce{W}}         & \multicolumn{2}{c}{\ce{V}}        & \multicolumn{2}{c}{\ce{Ru}}  \\
\midrule
\multirow{5}{*}{EML}
&$d$                      & \si{mm}  &  \num{6}         &  \num{12}       &  \num{6}         &  \num{12}       &  \num{6}        &  \num{12}       &  \num{6}         &  \num{12}   \\
&$I_s$                    & \si{A}   &  \num{177.9}     &  \num{169.1}    &  \num{532.6}     &  \num{496.2}    &  \num{310.1}    &  \num{286.7}    &  \num{420.7}     &  \num{399.9}     \\
&$u_\text{max}$           & \si{m/s} &  \num{0.738}     &  \num{1.150}    &  \num{2.17}      &  \num{3.05}     &  \num{1.46}     &  \num{2.15}     &  \num{1.46}      &  \num{2.18}      \\
&$\Rey$                   &          &  \num{7730}      &  \num{24052}    &  \num{31064}     &  \num{87208}    &  \num{11117}    &  \num{32736}    &  \num{15483}     &  \num{46168}     \\
&$T_\text{max}$           & \si{K}   &  \num{2046}      &  \num{2071}     &  \num{3979}      &  \num{3408}     &  \num{2791}     &  \num{2464}     &  \num{3266}      &  \num{3023}      \\
&$T_\text{min}$           & \si{K}   &  \num{2028}      &  \num{2050}     &  \num{3920}      &  \num{3376}     &  \num{2772}     &  \num{2447}     &  \num{3248}      &  \num{3008}      \\
\midrule                                                                                                                                                              
& $d$                     & \si{mm}  &  \num{2}         & \num{3}         &  \num{2}         & \num{3}         &  \num{2}        & \num{3}         &  \num{2}         & \num{3}   \\
        &$I_0$            & \si{W}   &  \num{11.5}      &  ---            &  \num{40}        &  ---            &  \num{18}       &  ---            &  \num{26}        & ---       \\
ADL     &$u_\text{max}\times 10^{-3}$& \si{m/s} & \num{6.61} &  ---       &  \num{9.52}      &  ---            &  \num{8.98}     &  ---            &  \num{3.91}      & ---       \\
(Ma)    &$\Rey$           &          &  \num{23}        &  ---            &  \num{45}        &  ---            &  \num{23}       &  ---            &  \num{14}        & ---       \\
        &$T_\text{max}$   & \si{K}   &  \num{2200}      &  ---            &  \num{3982}      &  ---            &  \num{2633}     &  ---            &  \num{2754}      & ---       \\
        &$T_\text{min}$   & \si{K}   &  \num{1937}      &  ---            &  \num{3721}      &  ---            &  \num{2440}     &  ---            &  \num{2614}      & ---       \\
\midrule                                                                                                                                                              
        &$\phi_\text{jet}$& \si{L/min}& \num{0.508}     &  \num{1.53}     &  \num{1.13}      &  \num{3.30}     &  \num{0.581}    &  \num{1.74}     &  \num{0.878}     &  \num{2.55}         \\
        &$u_\text{jet}$   & \si{m/s} &  \num{7.49}      &  \num{10.0}     &  \num{16.6}      &  \num{21.6}     &  \num{8.56}     &  \num{11.4}     &  \num{12.9}      &  \num{16.7}         \\
ADL     &$\Rey_\text{jet}$&          &  \num{1497}      &  \num{3006}     &  \num{3316}      &  \num{6484}     &  \num{1712}     &  \num{3419}     &  \num{2588}      &  \num{5010}         \\
(shear) &$u_\text{max}\times 10^{-1}$& \si{m/s}& \num{2.65} & \num{3.34}  &  \num{5.70}      &  \num{6.81}     &  \num{2.99}     &  \num{4.01}     &  \num{4.42}      &  \num{5.48}   \\
        &$\Rey$           &          &  \num{923}       &  \num{1751}     &  \num{2716}      &  \num{4866}     &  \num{760}      &  \num{1528}     &  \num{1557}      &  \num{2899}       \\
\midrule                                                                                                                                                              
\multirow{5}{*}{ESL}                                                                                                                                                  
&$d$                      & \si{mm}  &  \num{1.5}       &  \num{2.0}      &  \num{1.5}       &  \num{2.0}      &  \num{1.5}      &  \num{2.0}      &  \num{1.5}       &  \num{2.0}       \\
&$I_0$                    & \si{W}   &  \num{2}         &  \num{4}        &  \num{9}         &  \num{17}       &  \num{2}        &  \num{3}        &  \num{2.5}       &  \num{4}         \\
&$u_\text{max}\times 10^{-3}$ & \si{m/s} & \num{2.26}   &  \num{4.15}     &  \num{4.59}      &  \num{7.18}     &  \num{1.80}     &  \num{2.26}     &  \num{0.502}     &  \num{0.820}     \\
&$\Rey$                   &          &  \num{5.9}       &  \num{15}       &  \num{16}        &  \num{34}       &  \num{3.4}      &  \num{5.7}      &  \num{1.3}       &  \num{2.9}       \\
&$T_\text{max}$           & \si{K}   &  \num{2364}      &  \num{2294}     &  \num{3951}      &  \num{4236}     &  \num{2447}     &  \num{2422}     &  \num{2831}      &  \num{2892}      \\
&$T_\text{min}$           & \si{K}   &  \num{2276}      &  \num{2096}     &  \num{3818}      &  \num{3982}     &  \num{2397}     &  \num{2355}     &  \num{2805}      &  \num{2849}      \\
\bottomrule
\end{tabularx}
\end{table*}


\Cref{fig-ADL_convection} shows the
thermal and velocity fields for droplet internal convection for \TiSixFour of $d = \SI{2}{mm}$.
\Cref{fig-ADL_convection}(a) is the result for the Marangoni convection,
whereas \cref{fig-ADL_convection}(b) is for the shear-induced convection.
For the case of Marangoni convection,
nondimensional numbers of
$\La = \num{1.4e-1}$,
$\Ma = \num{4.1e4}$,
$\Bi = \num{4.6e-2}$, and
$\Pl = \num{3.1e1}$
are applied,
whereas for the case of shear-induced convection,
the applied nondimensional numbers are
$\Rey_\text{jet} = \num{1.5e3}$ and
$\nu_\ast = \num{1.7e1}$.
For both cases, the axisymmetic computaital domain is discretized by \num{21600} of meshes.
%
In \cref{fig-ADL_convection}(a),
the flow along the surface is driven from the hot spot to the cold spot by the Marangoni effect.
The maximum velocity is small as $u_\text{max} = \SI{6.61e-3}{m/s}$,
which corresponds to the Reynolds number $\Rey = \num{23}$.
For other materials of the same droplet size $d=\SI{2}{mm}$,
the Reynolds numbers are in the range \numrange{23}{45}, 
that is an order of magnitude smaller than those for the shear-induced flows described in the following section.
Therefore, the simulations of the Marangoni convection are only executed for the droplet size $d = \SI{2}{mm}$.

\Cref{fig-ADL_convection}(b) shows the velocity field for the shear-induced convection.
The flow is strongly driven 
where the shear stress $\tau_w$ takes a large value (\cref{fig-ADL_P_WSS}(b)).
The maximum velocity is $u_\text{max} = \SI{2.65e-1}{m/s}$,
which corresponds to the Reynolds number $\Rey = \num{923}$.
Similar calculations are conducted for other materials and droplet sizes,
and the Reynolds number is an order of magnitude $\Rey \sim 10^3$,
as summarized in \cref{tab-results}.

\subsection{Simulations for ESL}

\subsubsection{Configuration and procedure}
In the ESL model, the Marangoni effect caused by laser heating is the sole driving force for convection.
The configuration of the heating lasers is shown in \cref{fig-ESL_laser}, which is determined from the ISS-ELF.
Because of the non-axisymmetric layout of lasers, the three-dimensional calculation is necessary for thermal and velocity fields.
The power of laser $I_0$ was set such that the minimum temperature in the droplet
became larger than the melting point of the material.
Concerning the heat flux,
only the radiative heat loss was considered,
and the convective heat transfer was neglected.

\subsubsection{Droplet internal convection}
\Cref{fig-ESL_field} shows
temperature (contour) and velocity (vectors) fields
of the droplet internal flow in the  ESL system
for the case of \TiSixFour with $d = \SI{2}{mm}$.
The applied nondimensional numbers are
$\Pr = \num{6.6e-2}$,
$\Pl = \num{2.7e1}$,
$\Bi = \num{1.1e-3}$,
$\Ma = \num{4.2e4}$, and
$\La = \num{4.6e-2}$.
The three dimensional computaital domain is discretized by \num{108000} of hexahedral meshes.
The Marangoni effect, which drives the flow from the hot to the cold region along the surface, is shown.
The maximum velocity is $u_\text{max} = \SI{4.15e-3}{m/s}$,
which corresponds to Reynolds number $\Rey = \num{15}$.
Similar calculations were conducted for other materials and droplet sizes,
and the Reynolds number was an order of magnitude $\Rey \sim 10$,
as summarized in \cref{tab-results}.

\begin{figure*}
\centering
\begin{tabular}{lr}
\begin{minipage}[c]{0.40\textwidth}
\includegraphics[width=\textwidth]{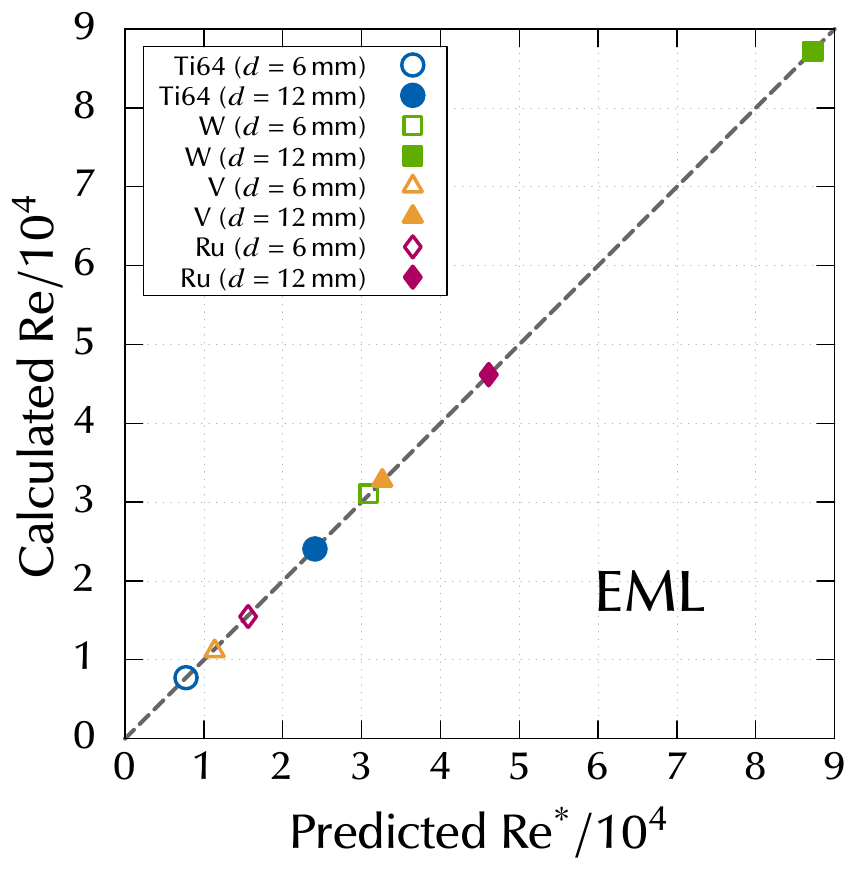}
\end{minipage}
 &
\begin{minipage}[c]{0.55\textwidth}
\small
\[
\begin{split}
 \Rey^\ast_\text{EML} &= a_m 
 \left(\frac{ \PR }{d_1} \right)^{m_1} 
 \left(\frac{ \Ga }{d_2} \right)^{m_2} 
 \left(\frac{ \Pm }{d_3} \right)^{m_3} 
 \left(\frac{ \Ma }{d_4} \right)^{m_4} \\
& \times
 \left(\frac{ \Ec }{d_5} \right)^{m_5}
 \left(\frac{ \Bi }{d_6} \right)^{m_6}
 \left(\frac{ \Pl }{d_7} \right)^{m_7}
 \left(\frac{ \Sp }{d_8} \right)^{m_8}
 \left(\frac{ \Mg }{d_9} \right)^{m_9}
\end{split}
\]
\centering
\sisetup{retain-zero-exponent=true}
\begin{tabularx}{0.7\textwidth}{cRR}
\toprule
$a_m$ & \multicolumn{2}{c}{ \num[round-mode=places,round-precision=3]{ 3.39364e+05}  }  \\
\midrule
$i$ & \multicolumn{1}{c}{$m_i$} & \multicolumn{1}{c}{$d_i$}  \\
\midrule
1     & \num[round-mode=places,round-precision=3]{-8.44868e-02}  &  \num[round-mode=places,round-precision=0]{1e-2}   \\
2     & \num[round-mode=places,round-precision=3]{ 6.46003e-01}  &  \num[round-mode=places,round-precision=0]{1e7 }   \\
3     & \num[round-mode=places,round-precision=3]{-8.45344e-09}  &  \num[round-mode=places,round-precision=0]{1e-7}   \\
4     & \num[round-mode=places,round-precision=3]{ 7.74434e-01}  &  \num[round-mode=places,round-precision=0]{1e5 }   \\
5     & \num[round-mode=places,round-precision=3]{-9.82028e-04}  &  \num[round-mode=places,round-precision=0]{1e14}   \\
6     & \num[round-mode=places,round-precision=3]{-9.92316e-01}  &  \num[round-mode=places,round-precision=0]{1e-4}   \\
7     & \num[round-mode=places,round-precision=3]{ 5.94993e-02}  &  \num[round-mode=places,round-precision=0]{1e1 }   \\
8     & \num[round-mode=places,round-precision=3]{-9.14696e-09}  &  \num[round-mode=places,round-precision=0]{1e2 }   \\
9     & \num[round-mode=places,round-precision=3]{ 2.65686e-02}  &  \num[round-mode=places,round-precision=0]{1e3 }   \\
\bottomrule
\end{tabularx}
\end{minipage}
\end{tabular}
\caption{\label{fig-resEML} 
Reynolds numbers of the droplet internal flow in the EML system.
The vertical axis represents $\Rey$ evaluated from the CFD result,
whereas the horizontal axis represents $\Rey^\ast$ predicted by the proposed formula of \cref{eq-pred_EML}.
The detailed coefficients and exponents are listed in the table on the right.
}
\vskip1em
\centering
\begin{tabular}{lr}
\begin{minipage}[c]{0.40\textwidth}
\includegraphics[width=\textwidth]{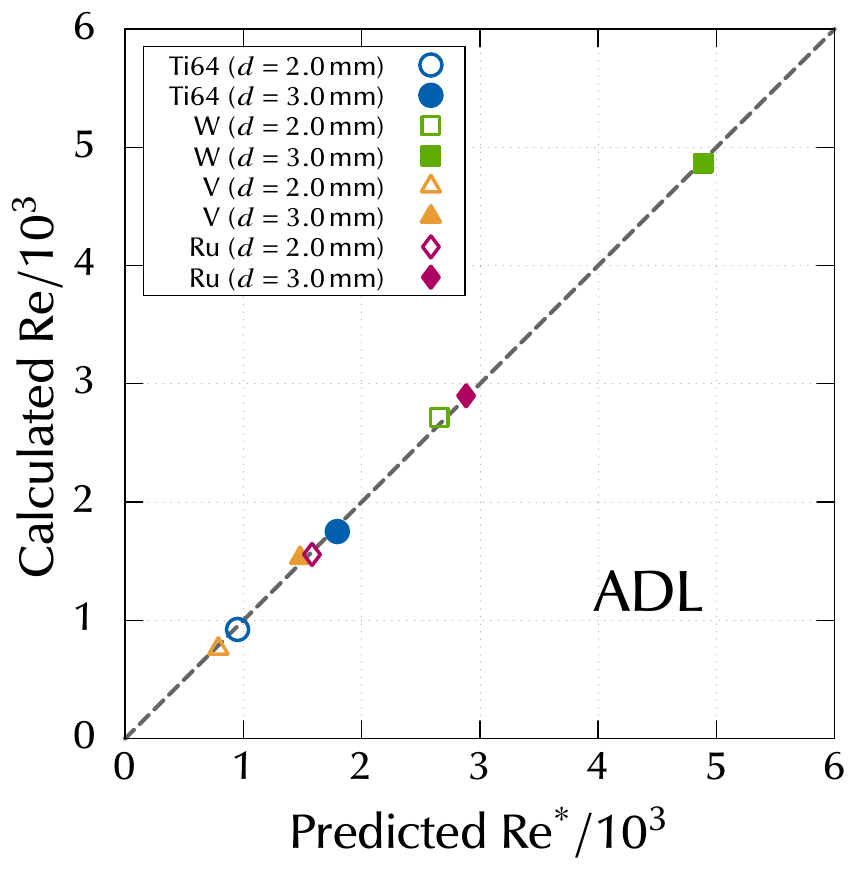}
\end{minipage}
 &
\begin{minipage}[c]{0.55\textwidth}
\small
\[
 \Rey^\ast_\text{ADL} = a_m
 \left(\frac{ \Rey_\text{jet}}{d_1}\right)^{m_1}
 \left(\frac{ \nu_\ast       }{d_2}\right)^{m_2}
\]
\centering
\begin{tabularx}{0.7\textwidth}{cRR}
\toprule
$a_m$  & \multicolumn{2}{c}{\num[round-mode=places,round-precision=3]{3.83146e2} } \\
\midrule
$i$ & \multicolumn{1}{c}{$m_i$} & \multicolumn{1}{c}{$d_i$}  \\
\midrule
1  & \num[round-mode=places,round-precision=3]{9.09899e-01}  &  \num[round-mode=places,round-precision=0]{1e3}  \\
2  & \num[round-mode=places,round-precision=3]{9.74379e-01}  &  \num[round-mode=places,round-precision=0]{1e1}  \\
\bottomrule
\end{tabularx}
\end{minipage}
\end{tabular}
\caption{\label{fig-resADL}
Reynolds numbers of the {\em shear-induced} droplet internal flow in the ADL system.
The vertical axis represents $\Rey$ evaluated from the CFD result,
whereas the horizontal axis represents $\Rey^\ast$ predicted by the proposed formula of \cref{eq-pred_ADL}.
The detailed coefficients and exponents are listed in the table on the right.
}
\vskip1em
\centering
\begin{tabular}{lr}
\begin{minipage}[c]{0.40\textwidth}
\includegraphics[width=\textwidth]{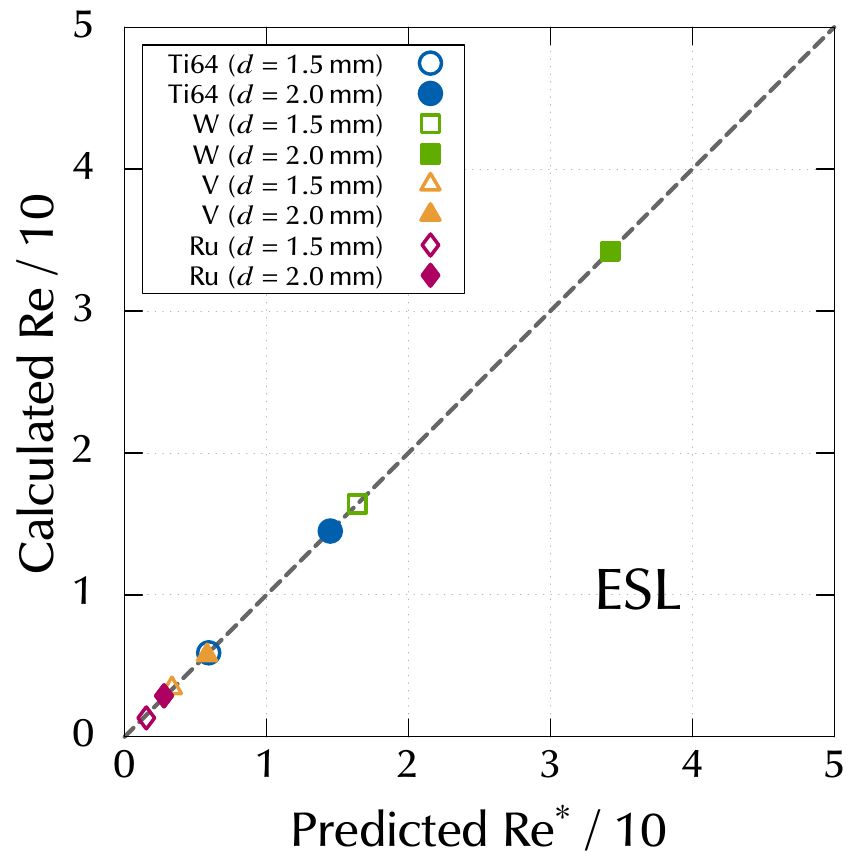}
\end{minipage}
 &
\begin{minipage}[c]{0.55\textwidth}
\small
\[
\begin{split}
 \Rey^\ast_\text{ESL} &= a_m 
 \left( \frac{ \PR }{d_1}\right)^{m_1} 
 \left( \frac{ \Bi }{d_2}\right)^{m_2}
 \left( \frac{ \Pl }{d_3}\right)^{m_3}
 \left( \frac{ \Ma }{d_4}\right)^{m_4}
 \left( \frac{ \La }{d_5}\right)^{m_5}
\end{split}
\]
\centering
\sisetup{retain-zero-exponent=true}
\begin{tabularx}{0.7\textwidth}{cRR}
\toprule
$a_m$ & \multicolumn{2}{c}{ \num[round-mode=places,round-precision=3]{8.03294e-01}} \\
\midrule
$i$ & \multicolumn{1}{c}{$m_i$} & \multicolumn{1}{c}{$d_i$}   \\
\midrule
1 & \num[round-mode=places,round-precision=3]{-2.54658e+00}  & \num[round-mode=places,round-precision=0]{1e-2}  \\
2 & \num[round-mode=places,round-precision=3]{-8.72366e-01}  & \num[round-mode=places,round-precision=0]{1e-3}  \\
3 & \num[round-mode=places,round-precision=3]{ 1.08780e+00}  & \num[round-mode=places,round-precision=0]{1e2}   \\
4 & \num[round-mode=places,round-precision=3]{ 3.41025e+00}  & \num[round-mode=places,round-precision=0]{1e4}   \\
5 & \num[round-mode=places,round-precision=3]{ 1.10606e+00}  & \num[round-mode=places,round-precision=0]{1e-3}  \\
\bottomrule
\end{tabularx}
\end{minipage}
\end{tabular}
\caption{\label{fig-resESL}
Reynolds numbers of the droplet internal flow in the ESL system.
The vertical axis represents $\Rey$ evaluated from the CFD result,
whereas the horizontal axis represents $\Rey^\ast$ predicted using the proposed formula of \cref{eq-pred_ESL}.
The detailed coefficients and exponents are listed in the table on the right.
}
\end{figure*}

\section{Surrogate models for prediction of internal flow}\label{sec-prediction}
The Reynolds number based on the maximum velocity is evaluated for all the simulation results obtained in this study, and they are summarized in \cref{tab-results}.
Although the models proposed in this study are formulated through assumptions and approximations, 
there are many related parameters, and obtaining the numerical results requires significant computational time.
If we want to know the Reynolds number for the material, 
which was not previously calculated, it is hard to interpolate from the results shown in \cref{tab-results}.

In this section, we propose simple surrogate formulas for predicting the Reynolds number of the droplet internal convection for the three levitation systems. 
The formulas are composed of combinations of nondimensional numbers that can be determined using the physical properties, system sizes, and driving conditions.
For the three levitation systems,
the formulas are written as follows:
\begin{align}
 \Rey^\ast_\text{EML} &= a_m 
 \left(\frac{ \PR }{d_1} \right)^{m_1} 
 \left(\frac{ \Ga }{d_2} \right)^{m_2} 
 \left(\frac{ \Pm }{d_3} \right)^{m_3} 
 \left(\frac{ \Ma }{d_4} \right)^{m_4} \notag \\
&\times
 \left(\frac{ \Ec }{d_5} \right)^{m_5}
 \left(\frac{ \Bi }{d_6} \right)^{m_6}
 \left(\frac{ \Pl }{d_7} \right)^{m_7}
 \left(\frac{ \Sp }{d_8} \right)^{m_8}
 \left(\frac{ \Mg }{d_9} \right)^{m_9}, \label{eq-pred_EML}   \\
 \Rey^\ast_\text{ADL} & = a_m
 \left(\frac{\Rey_\text{jet}}{d_1}\right)^{m_1}
 \left(\frac{\nu_\ast}{d_2}\right)^{m_2}, \label{eq-pred_ADL}
\\
 \Rey^\ast_\text{ESL} &= a_m 
 \left(\frac{ \PR }{d_1} \right)^{m_1} 
 \left(\frac{ \Bi }{d_2} \right)^{m_2}
 \left(\frac{ \Pl }{d_3} \right)^{m_3}
 \left(\frac{ \Ma }{d_4} \right)^{m_4}
 \left(\frac{ \La }{d_5} \right)^{m_5}, \label{eq-pred_ESL}
\end{align}
where the $\Rey^\ast$ is the predicted Reynolds number.
Denominators $d_i$ were selected as the orders of magnitude in the corresponding nondimensional number.
Coefficients $a_m$ and exponents $m_i$ were
determined from the numerical results shown in \cref{tab-results} to
minimize the following objective function:
\begin{equation}
 \mathcal{J} = \frac{1}{N} \sum_{j=1}^{N} \left( \text{Re}_j^\ast - \text{Re}_j^\text{CFD} \right)^2,  \label{eq-optJ}
\end{equation}
where $\Rey^\text{CFD}$ is Reynolds number calculated by the CFD.
Because the role of each nondimensional number, can be predicted regardless of whether it acts as a drive or suppresses the flow, 
some constraints are imposed on the optimization problem \cref{eq-optJ}.

For the EML system,
the Marangoni number $\Ma$ and Galilei number $\Ga$ can be regarded as driving factors.
Magnetic number $\Mg$ is also a driving factor because the intensity of the electromagnetic field is proportional to $\Mg$.
The shielding parameter $\Sp$ can be regarded as a suppressing factor from \cref{eq-solA2}.
The Prandtl number $\Pr$, the Magnetic Prandtl number $\Pm$, and the Eckert number $Ec$ can be regarded as suppressing factors from \cref{eq-nonDimMa,eq-nonDim_EM}.
From boundary condition \cref{eq-nonDimHeat}, 
the heat gain is proportional to the Biot number $\Bi$ and Laser power number $\La$ and inversely proportional to the Planck number $\Pl$.

For the ADL system, the surrogate model is constructed for shear-induced convection.
This model is simply composed of two nondimensional numbers: $\Rey_\text{jet}$ and $\nu_\ast$.
The jet Reynolds number $\Rey_\text{jet}$ is the driving factor.
From the shear stress boundary condition \cref{eq-stressBC_ADL} applied on the droplet surface, 
the viscosity ratio $\nu_\ast = \nu_\text{gas} / \nu$ can also be regarded as a driving factor.

For the ESL system, the convection is driven by the Marangoni effect,
and the temperature distribution is caused by laser heating.
Therefore, the Marangoni number $\Ma$ and the laser power number $\La$ are considered as driving factors.
The signs of contribution of the Biot number $\Bi$, Prandtl number $\Pr$, and Planck number $\Pl$ can be regarded as the same as those used in the EML system.

Based on the above discussion, the exponents $m_i$ in \cref{eq-pred_EML} 
corresponding to driving factors must be positive, whereas suppressing factors must be negative.
These constraints on the signs of the exponents $m_i$ are imposed.
The optimization problem is solved by the \texttt{L-BFGS-B} optimizer, 
and the identified values for $a_m$ and $m_i$ are listed on the right side of \cref{fig-resEML,fig-resADL,fig-resESL}.
The validity of the proposed surrogate formulas can be confirmed in the left side of \cref{fig-resEML,fig-resADL,fig-resESL}.

\section{Concluding remarks}
In this study, droplet internal flows were investigated for the EML, ADL, and ESL systems.
Simple mathematical models were formulated by assuming spherical shape of droplets with spatial symmetry.
Based on the formulated models, numerical simulations were conducted 
for several materials and droplet sizes, and the results were evaluated in terms of the Reynolds number 
based on the maximum velocity in the droplet. The order of magnitude of Reynolds numbers was evaluated as 
$\Rey \sim 10^4$ for EML, 
$\Rey \sim 10^3$ for ADL, and 
$\Rey \sim 10^1$ for ESL.
In the range of the present numerical simulations,
the order of levitation method for the same material was not changed.
Using the numerical results,
we proposed simple surrogate formulas that are used to predict the Reynolds number of flow internal droplets
using combinations of nondimensional numbers determined from the physical properties of a material and the driving conditions.
The proposed equations can also be used to predict the approximate Reynolds numbers
for materials other than those used in this study.

\section*{Acknowledgment}
This study was conducted as part of the \href{https://humans-in-space.jaxa.jp/kibouser/subject/science/70412.html}{Hetero-3D project}, supported by JAXA.
%
This study was partly supported by the Grant-in-Aid for Front Loading Research from the Advisory Committee for Space Utilization Research in ISAS/JAXA.
The author (SS) acknowledges the support from JSPS KAKENHI JP22K03909.   
Another author (SO) acknowledges the support from JSPS KAKENHI JP20H02453.  
The authors are grateful to 
B.Eng. Chihiro Hanada (Waseda University) and
Dr. Chihiro Koyama (JAXA) 
for their supports on this study.
%
The calculations shown in the present work were executed on 
the Fujitsu PRIMERGY CX400M1/CX2550M5 (Oakbridge-CX)
in the Information Technology Center, The University of Tokyo.






\begin{thebibliography}{999}

\bibitem[Lee \em{et~al.}(2021)Lee, Katamreddy, Cho, Lee, and Lee]{Lee2021}
J.~Lee, S.~Katamreddy, Y.~C. Cho, S.~Lee and G.~W. Lee: Containerless Materials
  Processing for Materials Science on Earth and in Space,
\newblock  Materials Processing Fundamentals 2021,
\newblock
\newblock Springer International Publishing (2021) 187,
\newblock DOI:
  {\href{https://doi.org/10.1007/978-3-030-65253-1_16}{\detokenize{10.1007/978-3-030-65253-1_16}}}.

\bibitem[Kuribayashi \em{et~al.}(2020)Kuribayashi, Shirasawa, Hayasaka,
  Shiratori, and Ozawa]{Kuribayashi2020}
K.~Kuribayashi, S.~Shirasawa, Y.~Hayasaka, S.~Shiratori and S.~Ozawa:
  Containerless processing of metastable multiferroic composite in Ln-(Mn,
  Fe)-O system (Ln: Lanthanide),
\newblock J. Am. Ceram. Soc., {\bfseries 103} (2020) 4822,
\newblock DOI:
  {\href{https://doi.org/10.1111/jace.17194}{\detokenize{10.1111/jace.17194}}}.

\bibitem[Hayasaka \em{et~al.}(2021)Hayasaka, Kuribayashi, Shiratori, and
  Ozawa]{Hayasaka2021}
Y.~Hayasaka, K.~Kuribayashi, S.~Shiratori and S.~Ozawa: Nucleation-Controlled
  Phase Selection in Rapid Solidification from Undercooled Melt of
  {DyMnO}{$_3$},
\newblock Mater. Trans., {\bfseries 62} (2021) 982,
\newblock DOI:
  {\href{https://doi.org/10.2320/matertrans.mt-m2021047}{\detokenize{10.2320/matertrans.mt-m2021047}}}.

\bibitem[Rayleigh(1879)]{Rayleigh1879}
L.~Rayleigh: On the capillary phenomena of jets,
\newblock Proc. R. Soc. London, {\bfseries 29} (1879) 71,
\newblock DOI:
  {\href{https://doi.org/10.1098/rspl.1879.0015}{\detokenize{10.1098/rspl.1879.0015}}}.

\bibitem[Bojarevics and Pericleous(2009)]{Bojarevics2009}
V.~Bojarevics and K.~Pericleous: Levitated droplet oscillations: effect of
  internal flow,
\newblock Magnetohydrodynamics, {\bfseries 45} (2009) 475,
\newblock DOI:
  {\href{https://doi.org/10.22364/mhd.45.3.22}{\detokenize{10.22364/mhd.45.3.22}}}.

\bibitem[Cummings and Blackburn(1991)]{Cummings1991}
D.~L. Cummings and D.~A. Blackburn: Oscillations of magnetically levitated
  aspherical droplets,
\newblock J. Fluid Mech., {\bfseries 224} (1991) 395,
\newblock DOI:
  {\href{https://doi.org/10.1017/s0022112091001817}{\detokenize{10.1017/s0022112091001817}}}.

\bibitem[Matson(2022)]{Matson2022}
D.~M. Matson: Metallurgy in Space,
\newblock In {\em Metallurgy in Space}; Springer International Publishing,
\newblock
\newblock chapter:  Influence of Convection on Phase Selection (2022) 299,
\newblock DOI:
  {\href{https://doi.org/10.1007/978-3-030-89784-0_14}{\detokenize{10.1007/978-3-030-89784-0_14}}}.

\bibitem[McCartney(1989)]{McCartney1989}
D.~G. McCartney: Grain refining of aluminium and its alloys using inoculants,
\newblock Int. Mater. Rev., {\bfseries 34} (1989) 247,
\newblock DOI:
  {\href{https://doi.org/10.1179/imr.1989.34.1.247}{\detokenize{10.1179/imr.1989.34.1.247}}}.

\bibitem[Tedman-Jones \em{et~al.}(2019)Tedman-Jones, McDonald, Bermingham,
  StJohn, and Dargusch]{TedmanJones2019}
S.~Tedman-Jones, S.~McDonald, M.~Bermingham, D.~StJohn and M.~Dargusch: A new
  approach to nuclei identification and grain refinement in titanium alloys,
\newblock J. Alloys Compd., {\bfseries 794} (2019) 268,
\newblock DOI:
  {\href{https://doi.org/10.1016/j.jallcom.2019.04.224}{\detokenize{10.1016/j.jallcom.2019.04.224}}}.

\bibitem[Watanabe \em{et~al.}(2020)Watanabe, Sato, Chiba, Sato, Sato, and
  Nakano]{Watanabe2020}
Y.~Watanabe, M.~Sato, T.~Chiba, H.~Sato, N.~Sato and S.~Nakano: 3D
  Visualization of Top Surface Structure and Pores of 3D Printed Ti-6Al-4V
  Samples Manufactured with {TiC} Heterogeneous Nucleation Site Particles,
\newblock Metall. Mater. Trans. A, {\bfseries 51} (2020) 1345,
\newblock DOI:
  {\href{https://doi.org/10.1007/s11661-019-05597-z}{\detokenize{10.1007/s11661-019-05597-z}}}.

\bibitem[Yamamoto \em{et~al.}(2019)Yamamoto, Date, Mori, Suzuki, Watanabe,
  Nakano, and Sato]{Yamamoto2019}
S.~Yamamoto, N.~Date, Y.~Mori, S.~Suzuki, Y.~Watanabe, S.~Nakano and N.~Sato:
  Effects of {TiC} Addition on Directionally Solidified Microstructure of
  Ti6Al4V,
\newblock Metall. Mater. Trans. A, {\bfseries 50} (2019) 3174,
\newblock DOI:
  {\href{https://doi.org/10.1007/s11661-019-05248-3}{\detokenize{10.1007/s11661-019-05248-3}}}.

\bibitem[Date \em{et~al.}(2021)Date, Yamamoto, Watanabe, Sato, Nakano, Sato,
  and Suzuki]{Date2021}
N.~Date, S.~Yamamoto, Y.~Watanabe, H.~Sato, S.~Nakano, N.~Sato and S.~Suzuki:
  Effects of Solidification Conditions on Grain Refinement Capacity of {TiC} in
  Directionally Solidified Ti6Al4V Alloy,
\newblock Metall. Mater. Trans. A, {\bfseries 52} (2021) 3609,
\newblock DOI:
  {\href{https://doi.org/10.1007/s11661-021-06333-2}{\detokenize{10.1007/s11661-021-06333-2}}}.

\bibitem[Tamaru \em{et~al.}(2018)Tamaru, Koyama, Saruwatari, Nakamura,
  Ishikawa, and Takada]{Tamaru2018}
H.~Tamaru, C.~Koyama, H.~Saruwatari, Y.~Nakamura, T.~Ishikawa and T.~Takada:
  Status of the Electrostatic Levitation Furnace ({ELF}) in the {ISS}-{KIBO},
\newblock Microgravity Sci. Tec., {\bfseries 30} (2018) 643,
\newblock DOI:
  {\href{https://doi.org/10.1007/s12217-018-9631-8}{\detokenize{10.1007/s12217-018-9631-8}}}.

\bibitem[Ishikawa \em{et~al.}(2022)Ishikawa, Koyama, Oda, Saruwatari, and
  Paradis]{Ishikawa2022}
T.~Ishikawa, C.~Koyama, H.~Oda, H.~Saruwatari and P.-F. Paradis: Status of the
  Electrostatic Levitation Furnace in the ISS -Surface Tension and Viscosity
  Measurements,
\newblock Int. J. Microgravity Sci. Appl., {\bfseries 12} (2022) 390101,
\newblock DOI:
  {\href{https://doi.org/10.15011/jasma.39.390101}{\detokenize{10.15011/jasma.39.390101}}}.

\bibitem[Hanada \em{et~al.}(2023)Hanada, Aoki, Ueda, Kadoi, Mabuchi, Yoneda,
  Yamada, Sato, Watanabe, Harada, Ozawa, Nakano, Koyama, Oda, Ishikawa,
  Watanabe, Shimaoka, and Suzuki]{Hanada2023}
C.~Hanada, H.~Aoki, Y.~Ueda, K.~Kadoi, Y.~Mabuchi, K.~Yoneda, M.~Yamada,
  H.~Sato, Y.~Watanabe, Y.~Harada, S.~Ozawa, S.~Nakano, C.~Koyama, H.~Oda,
  T.~Ishikawa, Y.~Watanabe, T.~Shimaoka and S.~Suzuki: Suppression of bubble
  formation in levitated molten samples of {Ti6Al4V} with {TiC} for {\em
  Hetero-3D} at the {International Space Station (ISS)},
\newblock Int. J. Microgravity Sci. Appl., {\bfseries 40} (2023) 400301,
\newblock DOI:
  {\href{https://doi.org/10.15011/jasma.40.400301}{\detokenize{10.15011/jasma.40.400301}}}.

\bibitem[Bojarevics \em{et~al.}(2000)Bojarevics, Pericleous, and
  Cross]{Bojarevics2000}
V.~Bojarevics, K.~Pericleous and M.~Cross: Modeling the dynamics of magnetic
  semilevitation melting,
\newblock Metall. Mater. Trans. B, {\bfseries 31} (2000) 179,
\newblock DOI:
  {\href{https://doi.org/10.1007/s11663-000-0143-7}{\detokenize{10.1007/s11663-000-0143-7}}}.

\bibitem[Bojarevics and Pericleous(2003)]{Bojarevics2003}
V.~Bojarevics and K.~Pericleous: Modelling Electromagnetically Levitated Liquid
  Droplet Oscillations,
\newblock {ISIJ} International, {\bfseries 43} (2003) 890,
\newblock DOI:
  {\href{https://doi.org/10.2355/isijinternational.43.890}{\detokenize{10.2355/isijinternational.43.890}}}.

\bibitem[Berry \em{et~al.}(2000)Berry, Hyers, Abedian, and Racz]{Berry2000}
S.~Berry, R.~W. Hyers, B.~Abedian and L.~M. Racz: Modeling of turbulent flow in
  electromagnetically levitated metal droplets,
\newblock Metall. Mater. Trans. B, {\bfseries 31} (2000) 171,
\newblock DOI:
  {\href{https://doi.org/10.1007/s11663-000-0142-8}{\detokenize{10.1007/s11663-000-0142-8}}}.

\bibitem[Hyers \em{et~al.}(2003)Hyers, Trapaga, and Abedian]{Hyers2003}
R.~W. Hyers, G.~Trapaga and B.~Abedian: Laminar-turbulent transition in an
  electromagnetically levitated droplet,
\newblock Metall. Mater. Trans. B, {\bfseries 34} (2003) 29,
\newblock DOI:
  {\href{https://doi.org/10.1007/s11663-003-0052-7}{\detokenize{10.1007/s11663-003-0052-7}}}.

\bibitem[Tsukada \em{et~al.}(2009)Tsukada, ichi Sugioka, Tsutsumino, Fukuyama,
  and Kobatake]{Tsukada2009}
T.~Tsukada, K.~i.~Sugioka, T.~Tsutsumino, H.~Fukuyama and H.~Kobatake: Effect
  of static magnetic field on a thermal conductivity measurement of a molten
  droplet using an electromagnetic levitation technique,
\newblock Int. J. Heat Mass Transfer, {\bfseries 52} (2009) 5152,
\newblock DOI:
  {\href{https://doi.org/10.1016/j.ijheatmasstransfer.2009.04.020}{\detokenize{10.1016/j.ijheatmasstransfer.2009.04.020}}}.

\bibitem[Spitans \em{et~al.}(2013)Spitans, Jakovics, Baake, and
  Nacke]{Spitans2013}
S.~Spitans, A.~Jakovics, E.~Baake and B.~Nacke: Numerical Modeling of Free
  Surface Dynamics of Melt in an Alternate Electromagnetic Field: Part I.
  Implementation and Verification of Model,
\newblock Metall. Mater. Trans. B, {\bfseries 44} (2013) 593,
\newblock DOI:
  {\href{https://doi.org/10.1007/s11663-013-9809-9}{\detokenize{10.1007/s11663-013-9809-9}}}.

\bibitem[Spitans \em{et~al.}(2016)Spitans, Baake, Nacke, and
  Jakovics]{Spitans2016}
S.~Spitans, E.~Baake, B.~Nacke and A.~Jakovics: Numerical Modeling of Free
  Surface Dynamics of Melt in an Alternate Electromagnetic Field. Part {II}:
  Conventional Electromagnetic Levitation,
\newblock Metall. Mater. Trans. B, {\bfseries 47} (2016) 522,
\newblock DOI:
  {\href{https://doi.org/10.1007/s11663-015-0515-7}{\detokenize{10.1007/s11663-015-0515-7}}}.

\bibitem[Guo \em{et~al.}(2019)Guo, Manickam, Yu, Villanueva, and Ma]{Guo2019}
Q.~Guo, L.~Manickam, P.~Yu, W.~Villanueva and W.~Ma: A design study on an
  aerodynamic levitation system for droplet preparation in steam explosion
  experiment,
\newblock  Proc. 27th Int. Conf. on Nuclear Engineering (ICONE27),
\newblock
\newblock Japan Society of Mechanical Engineers (2019) 2366,
\newblock DOI:
  {\href{https://doi.org/10.1299/jsmeicone.2019.27.2366}{\detokenize{10.1299/jsmeicone.2019.27.2366}}}.

\bibitem[Song and Li(2000)]{Song2000}
S.~Song and B.~Li: Free surface profiles and thermal convection in
  electrostatically levitated droplets,
\newblock Int. J. Heat Mass Transfer, {\bfseries 43} (2000) 3589,
\newblock DOI:
  {\href{https://doi.org/10.1016/s0017-9310(00)00004-1}{\detokenize{10.1016/s0017-9310(00)00004-1}}}.

\bibitem[Huo and Li(2004)]{Huo2004}
Y.~Huo and B.~Li: Three-dimensional Marangoni convection in electrostatically
  positioned droplets under microgravity,
\newblock Int. J. Heat Mass Transfer, {\bfseries 47} (2004) 3533,
\newblock DOI:
  {\href{https://doi.org/10.1016/j.ijheatmasstransfer.2004.01.021}{\detokenize{10.1016/j.ijheatmasstransfer.2004.01.021}}}.

\bibitem[Hyers \em{et~al.}(2004)Hyers, Matson, Kelton, and Rogers]{Hyers2004}
R.~W. Hyers, D.~M. Matson, K.~F. Kelton and J.~R. Rogers: Convection in
  Containerless Processing,
\newblock Annals of the New York Academy of Sciences, {\bfseries 1027} (2004)
  474,
\newblock DOI:
  {\href{https://doi.org/10.1196/annals.1324.038}{\detokenize{10.1196/annals.1324.038}}}.

\bibitem[Hyers(2005)]{Hyers2005}
R.~W. Hyers: Fluid flow effects in levitated droplets,
\newblock Meas. Sci. Technol., {\bfseries 16} (2005) 394,
\newblock DOI:
  {\href{https://doi.org/10.1088/0957-0233/16/2/010}{\detokenize{10.1088/0957-0233/16/2/010}}}.

\bibitem[Gao \em{et~al.}(2016)Gao, Shi, Li, Yang, Zhang, McLean, and
  Chattopadhyay]{Gao2016}
L.~Gao, Z.~Shi, D.~Li, Y.~Yang, G.~Zhang, A.~McLean and K.~Chattopadhyay:
  Dimensionless Analysis and Mathematical Modeling of Electromagnetic
  Levitation (EML) of Metals,
\newblock Metall. Mater. Trans. B, {\bfseries 47} (2016) 67,
\newblock DOI:
  {\href{https://doi.org/10.1007/s11663-015-0457-0}{\detokenize{10.1007/s11663-015-0457-0}}}.

\bibitem[Xiao \em{et~al.}(2019)Xiao, Lee, Hyers, and Matson]{Xiao2019}
X.~Xiao, J.~Lee, R.~W. Hyers and D.~M. Matson: npj Microgravity, {\bfseries 5}
  (2019),
\newblock DOI:
  {\href{https://doi.org/10.1038/s41526-019-0067-2}{\detokenize{10.1038/s41526-019-0067-2}}}.

\bibitem[Baker \em{et~al.}(2020)Baker, Nawer, Xiao, and Matson]{Baker2020}
E.~B. Baker, J.~Nawer, X.~Xiao and D.~M. Matson: npj Microgravity, {\bfseries
  6} (2020),
\newblock DOI:
  {\href{https://doi.org/10.1038/s41526-020-0099-7}{\detokenize{10.1038/s41526-020-0099-7}}}.

\bibitem[Smythe(1968)]{Smythe1968}
W.~R. Smythe: Static and Dynamic Electricity, 3rd ed.,
\newblock McGraw-Hill, (1968).

\bibitem[Li(1993)]{Li1993}
B.~Li: The magnetothermal phenomena in electromagnetic levitation processes,
\newblock Int. J. Eng. Sci., {\bfseries 31} (1993) 201,
\newblock DOI:
  {\href{https://doi.org/10.1016/0020-7225(93)90034-r}{\detokenize{10.1016/0020-7225(93)90034-r}}}.

\bibitem[Ishikawa \em{et~al.}(2011)Ishikawa, Okada, Paradis, and
  Watanabe]{Ishikawa_2011}
T.~Ishikawa, J.~T. Okada, P.-F. Paradis and Y.~Watanabe: Thermophysical
  Property Measurements of High Temperature Melts Using an Electrostatic
  Levitation Method,
\newblock Jpn. J. Appl. Phys., {\bfseries 50} (2011) 11RD03,
\newblock DOI:
  {\href{https://doi.org/10.1143/JJAP.50.11RD03}{\detokenize{10.1143/JJAP.50.11RD03}}}.

\bibitem[Holfelder and Witte(2020)]{Holfelder_2020}
P.~Holfelder and A.~Witte: Simulation-assisted analysis of microstructural
  evolution of Ti{\textendash}6Al{\textendash}4V during laser powder bed
  fusion,
\newblock Prog. Addit. Manuf., {\bfseries 5} (2020) 237,
\newblock DOI:
  {\href{https://doi.org/10.1007/s40964-020-00114-w}{\detokenize{10.1007/s40964-020-00114-w}}}.

\bibitem[Pottlacher(1999)]{Pottlacher_1999}
G.~Pottlacher: Thermal conductivity of pulse-heated liquid metals at melting
  and in the liquid phase,
\newblock J. Non-Cryst. Solids, {\bfseries 250-252} (1999) 177,
\newblock DOI:
  {\href{https://doi.org/10.1016/S0022-3093(99)00116-7}{\detokenize{10.1016/S0022-3093(99)00116-7}}}.

\bibitem[Pottlacher \em{et~al.}(2007)Pottlacher, H\"{u}pf, Wilthan, and
  Cagran]{Pottlacher2007}
G.~Pottlacher, T.~H\"{u}pf, B.~Wilthan and C.~Cagran: Thermophysical data of
  liquid vanadium,
\newblock Thermochim. Acta, {\bfseries 461} (2007) 88,
\newblock DOI:
  {\href{https://doi.org/10.1016/j.tca.2006.12.010}{\detokenize{10.1016/j.tca.2006.12.010}}}.

\bibitem[Paradis \em{et~al.}(2004)Paradis, Ishikawa, and Yoda]{Paradis2004}
P.-F. Paradis, T.~Ishikawa and S.~Yoda: Thermophysical properties of liquid and
  supercooled ruthenium measured by noncontact methods,
\newblock J. Mater. Res., {\bfseries 19} (2004) 590,
\newblock DOI:
  {\href{https://doi.org/10.1557/jmr.2004.19.2.590}{\detokenize{10.1557/jmr.2004.19.2.590}}}.

\bibitem[Mohr \em{et~al.}(2020)Mohr, Wunderlich, Novakovic, Ricci, and
  Fecht]{Mohr_2020}
M.~Mohr, R.~Wunderlich, R.~Novakovic, E.~Ricci and H.-J. Fecht: Precise
  Measurements of Thermophysical Properties of Liquid
  Ti{\textendash}6Al{\textendash}4V (Ti64) Alloy On Board the International
  Space Station,
\newblock Adv. Eng. Mater., {\bfseries 22} (2020) 2000169,
\newblock DOI:
  {\href{https://doi.org/10.1002/adem.202000169}{\detokenize{10.1002/adem.202000169}}}.

\bibitem[Ho \em{et~al.}(1972)Ho, Powell, and Liley]{Ho1972}
C.~Y. Ho, R.~W. Powell and P.~E. Liley: Thermal Conductivity of the Elements,
\newblock J. Phys. Chem. Ref. Data, {\bfseries 1} (1972) 279,
\newblock DOI:
  {\href{https://doi.org/10.1063/1.3253100}{\detokenize{10.1063/1.3253100}}}.

\bibitem[Joshi \em{et~al.}(2017)Joshi, Zinzala, Nirmal, and Fuse]{Joshi_2017}
R.~Joshi, G.~Zinzala, N.~Nirmal and K.~Fuse: Multi-Response Optimization of
  {EDM} for Ti-6Al-4V Using Taguchi - Grey Relational Analysis,
\newblock Solid State Phenomena, {\bfseries 266} (2017) 43,
\newblock DOI:
  {\href{https://doi.org/10.4028/www.scientific.net/SSP.266.43}{\detokenize{10.4028/www.scientific.net/SSP.266.43}}}.

\bibitem[White and Minges(1997)]{White1997}
G.~K. White and M.~L. Minges: Thermophysical properties of some key solids: An
  update,
\newblock Int. J. Thermophys., {\bfseries 18} (1997) 1269,
\newblock DOI:
  {\href{https://doi.org/10.1007/bf02575261}{\detokenize{10.1007/bf02575261}}}.

\bibitem[Desai \em{et~al.}(1984)Desai, James, and Ho]{Desai_1984}
P.~D. Desai, H.~M. James and C.~Y. Ho: Electrical Resistivity of Vanadium and
  Zirconium,
\newblock J. Phys. Chem. Ref. Data, {\bfseries 13} (1984) 1097,
\newblock DOI:
  {\href{https://doi.org/10.1063/1.555724}{\detokenize{10.1063/1.555724}}}.

\bibitem[Arblaster(2016)]{Arblaster_2016}
J.~W. Arblaster: Selected Electrical Resistivity Values for the Platinum Group
  of Metals Part {III}: Ruthenium and Osmium,
\newblock Johnson Matthey Technology Review, {\bfseries 60} (2016) 179,
\newblock DOI:
  {\href{https://doi.org/10.1595/205651316X691618}{\detokenize{10.1595/205651316X691618}}}.

\bibitem[Milo{\v{s}}evi{\'{c}} and Aleksi{\'{c}}(2012)]{Milo_evi__2012}
N.~Milo{\v{s}}evi{\'{c}} and I.~Aleksi{\'{c}}: Thermophysical properties of
  solid phase Ti-6Al-4V alloy over a wide temperature range,
\newblock Int. J. Mater. Res., {\bfseries 103} (2012) 707,
\newblock DOI:
  {\href{https://doi.org/10.3139/146.110678}{\detokenize{10.3139/146.110678}}}.

\bibitem[Allen \em{et~al.}(1960)Allen, Glasier, and Jordan]{Allen_1960}
R.~D. Allen, L.~F. Glasier and P.~L. Jordan: Spectral Emissivity, Total
  Emissivity, and Thermal Conductivity of Molybdenum, Tantalum, and Tungsten
  above 2300{\textdegree}K,
\newblock J. Appl. Phys., {\bfseries 31} (1960) 1382,
\newblock DOI:
  {\href{https://doi.org/10.1063/1.1735847}{\detokenize{10.1063/1.1735847}}}.

\bibitem[Cezairliyan \em{et~al.}(1974)Cezairliyan, Righini, and
  McClure]{Cezairliyan_1974}
A.~Cezairliyan, F.~Righini and J.~McClure: Simultaneous measurements of heat
  capacity, electrical resistivity, and hemispherical total emittance by a
  pulse heating technique: Vanadium, 1500 to 2100 K,
\newblock J. Res. Natl. Bur. Stand., A Phys. Chem., {\bfseries 78A} (1974) 143,
\newblock DOI:
  {\href{https://doi.org/10.6028/jres.078A.010}{\detokenize{10.6028/jres.078A.010}}}.

\bibitem[Milo{\v{s}}evi{\'{c}} and Nikoli{\'{c}}(2015)]{Milo_evi__2015}
N.~Milo{\v{s}}evi{\'{c}} and I.~Nikoli{\'{c}}: Thermophysical properties of
  solid phase ruthenium measured by the pulse calorimetry technique over a wide
  temperature range,
\newblock Int. J. Mater. Res., {\bfseries 106} (2015) 361,
\newblock DOI:
  {\href{https://doi.org/10.3139/146.111192}{\detokenize{10.3139/146.111192}}}.

\bibitem[Ozawa \em{et~al.}(2021)Ozawa, Nagasaka, Itakura, Sugisawa, and
  Seimiya]{Ozawa_2021}
S.~Ozawa, Y.~Nagasaka, M.~Itakura, K.~Sugisawa and Y.~Seimiya: Influence of
  oxygen adsorption from atmosphere on surface tension of liquid silicon,
\newblock J. Appl. Phys., {\bfseries 131} (2021) 129902,
\newblock DOI:
  {\href{https://doi.org/10.1063/5.0062062}{\detokenize{10.1063/5.0062062}}}.

\bibitem[Incropera \em{et~al.}(2007)Incropera, Dewitt, Bergman, and
  Lavine]{Incropera2007}
F.~P. Incropera, D.~P. Dewitt, T.~L. Bergman and A.~S. Lavine: Fundamentals of
  Heat and Mass Transfer, 6th ed.,
\newblock John Wiley \& Sons, (2007).

\end{thebibliography}


\section*{Nomenclature}
\small
\begin{tabularx}{\textwidth}{cclccl}
\toprule
Symbol                      & Units                &  Description                                & Symbol                       & Units                &  Description                                  \\
\midrule
$\bol{A}$                   & \si{N/A}             &  Magnetic potential                         &  $u_\text{jet}$              & \si{m/s}             &  Jet velocity                               \\
$\bol{B}$                   & \si{N/(A m)}         &  Magnetic flux density                      &  $\bol{U}$                   & N.D.                 &  Nondimensional velocity                    \\
$B_0$                       & \si{N/(A m)}         &  Scale for magnetic flux density            &  $W(x)$                      & \si{1/m^2}           &  Spatial distribution of laser spot         \\
$\widehat{\bol{B}}$         & N.D.                 &  Nondimensional magnetic flux density       &  $x,y,z$                     & \si{m}               &  Components of coordinates                  \\
$c_p$                       & \si{J/(kg K)}        &  Specific heat                              &  $\bol{x}$                   & \si{m}               &  Spatial coordinates                        \\
$C_D$                       & N.D.                 &  Drag coefficient                           &  $\bol{X}$                   & N.D.                 &  Nondimensional coordinates                 \\
$d$                         & \si{m}               &  Droplet diameter                           &  $z_c$                       & \si{m}               &  Axial position of droplet center           \\
$d_\text{noz}$              & \si{m}               &  Nozzle outlet diameter                     &  $\alpha$                    & \si{m^2/s}           &  Thermal diffusivity                        \\
$\bol{D}$                   & \si{s^{-1}}          &  Strain rate tensor                         &  $\varepsilon$               & N.D.                 &  Emissivity                                 \\
$\bol{e}_g$                 & N.D.                 &  Unit vector along gravity direction        &  $\eta$                      & \si{1/m}             &  Attenuation coefficient                    \\
$\bol{e}_z$                 & N.D.                 &  Unit vector along axial direction          &  $\Theta$                    & N.D                  &  Nondimensional temperature                 \\
$\bol{f}_m$                 & \si{N/m^3}           &  Lorentz force per unit volume              &  $\lambda$                   & \si{W/(m K)}         &  Thermal conductivity                       \\
$F_m$                       & \si{N}               &  Levitation force                           &  $\mu$                       & \si{Pa s}            &  Viscosity                                  \\
$\bol{g}$                   & \si{m/s^2}           &  Gravitational acceleration                 &  $\mu_0$                     & \si{N/A^2}           &  Permeability of free space                 \\
$h$                         & \si{W/(m^2 K)}       &  Heat transfer coefficient                  &  $\nu$                       & \si{m^2/s}           &  Kinematic  viscosity                       \\
$i$                         & N.D.                 &  Imaginary unit                             &  $\nu_\ast$                  & N.D.                 &  Viscosity ratio                            \\
$I_0$                       & \si{W}               &  Power of the laser heat source             &  $\rho$                      & \si{kg/m^3}          &  Density                                    \\
$I_s$                       & \si{A}               &  Electrical current amplitude               &  $\sigma_e$                  & \si{S/m}             &  Electrical conductivity                    \\
$\bol{J}$                   & \si{A/m^2}           &  Electric current density                   &  $\sigma_\text{SB}$          & \si{W/(m^2 K^4)}     &  Stefan Boltzmann constant                  \\
$\widehat{\bol{J}}$         & N.D.                 &  Nondimensional current density             &  $\sigma_T$                  & \si{N/(m K)}         &  Temperature coefficient of surface tension \\
$\mathcal{J}$               & N.D.                 &  Objective function                         &  $\tau$                      & N.D.                 &  Nondimensional time                        \\
$L$                         & \si{m}               &  Representative length                      &  $\tau_w$                    & \si{Pa}              &  Surface shear stress                       \\
$\bol{n}$                   & N.D.                 &  Unit normal vector                         &  $\phi_\text{jet}$           & \si{m^3/s}           &  Volumetric flow rate                       \\
$p$                         & \si{Pa}              &  Pressure                                   &  $\varphi$                   & \si{rad}             &  Azimuthal angle                            \\
$P$                         & N.D                  &  Nondimensional pressure                    &  $\omega$                    & \si{rad/s}           &  Angular frequency of electric current      \\
$p_d$                       & \si{Pa}              &  Dynamic pressure                           &  $\Bi$                       & N.D.                 &  Biot number                                \\
$q_m$                       & \si{W/m^3}           &  Joule heat                                 &  $\Ec$                       & N.D.                 &  Eckert number                              \\
$r$                         & \si{m}               &  Radial coordinate                          &  $\Ga$                       & N.D.                 &  Galilei number                             \\
$R_0$                       & \si{m}               &  Droplet radius                             &  $\Pl$                       & N.D.                 &  Planck number                              \\
$R_L$                       & \si{m}               &  Radius of laser spot                       &  $\La$                       & N.D.                 &  Laser power number                         \\
$R_s, \theta_s$             & \si{m}, \si{rad}     &  Coil position in spherical coordinate      &  $\Ma$                       & N.D.                 &  Marangoni number                           \\
$s$                         & \si{m}               &  Distance from the axis of laser spot       &  $\Mg$                       & N.D.                 &  Magnetic number                            \\
$t$                         & \si{s}               &  Time                                       &  $\Nu$                       & N.D.                 &  Nusselt number                             \\
$T$                         & \si{K}               &  Temperature                                &  $\Pm$                       & N.D.                 &  Magnetic Prandtl number                    \\
$T_a$                       & \si{K}               &  Ambient temperature                        &  $\PR$                       & N.D.                 &  Prandtl number                             \\
$T_\ast$                    & \si{K}               &  Melting point                              &  $\Rey$                      & N.D.                 &  Reynolds number                            \\
$\bol{u}$                   & \si{m/s}             &  Velocity                                   &  $\Rey_\text{jet}$           & N.D.                 &  Jet Reynolds number                        \\
$u_\text{max}$              & \si{m/s}             &  Maximum velocity                           &  $\Sp$                       & N.D.                 &  Shielding parameter                        \\
\bottomrule
\end{tabularx}

\end{document}